\documentclass[aps,reprint,superscriptaddress,longbibliography]{revtex4-1}
\usepackage{times,graphicx,jab}
\usepackage{verbatim}
\usepackage{color}
\usepackage[T1]{fontenc}
\usepackage{mathtools}

\newcommand{\beq}{\begin{equation}}
\newcommand{\eeq}{\end{equation}}

\newcommand{\vA}{v_{\mathrm{A}}}
\newcommand{\dd}{\partial}
\newcommand{\vz}{\boldsymbol{z}}

\newcommand{\ddt}{\dd_t}

\newcommand{\ddz}{\dd_z}

\newcommand{\nap}{\nabla_\perp}

\newcommand{\hvz}{\hat{\vz}}
\newcommand{\brak}[2]{\left\{#1,#2\right\}}

\def\date#1{}

\begin{document}
\title{Strong Nonlinear Alfv\'en Wave Interactions in a Laboratory Plasma}
\date{\today}
\author{C.~H.~K.~Chen}
\affiliation{Department of Physics and Astronomy, Queen Mary University of London, London E1 4NS, UK}
\email{christopher.chen@qmul.ac.uk}
\author{S. Dorfman}
\affiliation{Space Science Institute, Boulder, CO 80301, USA}
\affiliation{Department of Physics and Astronomy, University of California, Los Angeles, Los Angeles, CA 90095, USA}
\author{S. Boldyrev}
\affiliation{Department of Physics, University of Wisconsin--Madison, Madison, WI 53706, USA}
\affiliation{Space Science Institute, Boulder, CO 80301, USA}
\author{L. Franci}
\affiliation{School of Engineering, Physics and Mathematics, Northumbria University, Newcastle upon Tyne, NE1 8ST, UK}
\affiliation{Department of Physics and Astronomy, Queen Mary University of London, London E1 4NS, UK}
\author{A. Mallet}
\affiliation{Space Sciences Laboratory, University of California, Berkeley, Berkeley, CA 94720, USA}
\author{M. Abler}
\affiliation{Space Science Institute, Boulder, CO 80301, USA}
\affiliation{Department of Physics and Astronomy, University of California, Los Angeles, Los Angeles, CA 90095, USA}
\author{S. Vincena}
\affiliation{Department of Physics and Astronomy, University of California, Los Angeles, Los Angeles, CA 90095, USA}
\author{S. Greess}
\affiliation{Department of Physics and Astronomy, Queen Mary University of London, London E1 4NS, UK}
\author{T. A. Carter}
\affiliation{Department of Physics and Astronomy, University of California, Los Angeles, Los Angeles, CA 90095, USA}
\affiliation{Fusion Energy Division, Oak Ridge National Laboratory, Oak Ridge, TN 37831, USA}
\begin{abstract}
Alfv\'en waves and their nonlinear interactions are ubiquitous in space and astrophysical plasmas, and are thought to play important roles in the dynamics of these systems, yet their nature remains to be fully understood. We describe experiments performed on the Large Plasma Device to study the nature of counter- and co-propagating wave interactions relevant to strong Alfv\'enic turbulence. Both interactions were found to produce a broad spectrum of nonlinear modes as a result of a dominant quadratic nonlinearity. The counter-propagating interaction can be explained through the standard reduced MHD nonlinearity, and the co-propagating interaction can be explained through a recently-proposed model that includes second-order nonlinear terms from Hall MHD that dominate at large imbalance and scale with the ion inertial length. The predictions of the latter model were tested in both the experiment and in 3D hybrid simulations, where the nonlinear mode growth rate and ion inertial scale dependence were found to be consistent. Finally, at the obtained interaction strengths, energy was seen to be transferred to progressively smaller perpendicular scales, consistent with a local cascade, although not a state of fully-developed turbulence. These results reveal and verify the mechanisms occurring in balanced and imbalanced turbulence (as well as other Alfv\'enic nonlinear processes), and represent an important step towards the generation of controlled Alfv\'enic turbulence in the laboratory.
\end{abstract}
\maketitle

Alfv\'en waves \citep{alfven42} are a fundamental plasma mode that are widespread in both astrophysical and laboratory plasmas. In most of these systems, they interact nonlinearly, producing broadband turbulence, complex structuring of the plasma, and influencing the system dynamics. For example, Alfv\'enic turbulence is thought to play a key role in coronal heating and solar wind acceleration \citep{cranmer19,chen22}, accretion disk transport \citep{balbus98}, star formation \citep{mckee07}, galactic dynamo \citep{kulsrud08}, and galaxy cluster heating \citep{zhuravleva14}. The solar wind provides unparalleled access to in situ measurements \citep{bruno13,alexandrova13a,chen16b,wilson21}, enabling detailed comparisons with theoretical models of Alfv\'enic turbulence \citep{schekochihin22}, and while a large number of these features are in agreement, open questions still remain. Laboratory measurements provide a complementary approach to understand the underlying physics of nonlinear Alfv\'en wave interactions and turbulence.

The theory of Alfv\'enic turbulence began with the work of Iroshnikov \citep{iroshnikov63t} and Kraichnan \citep{kraichnan65} who developed a model of weak isotropic turbulence based on the fact that in homogenous magnetohydrodynamics (MHD) Alfv\'en waves only interact nonlinearly if they are counter-propagating. This is because the reduced MHD nonlinear term $\delta\mathbf{z}^+\cdot\nabla\delta\mathbf{z}^-$ is only non-zero when both fluctuating Elsasser \citep{elsasser50} variables $\delta\mathbf{z}^\pm=\delta\mathbf{u}\pm\delta\mathbf{B}/\sqrt{\mu_0\rho}$ (where $\mathbf{u}$ is the velocity, $\mathbf{B}$ is the magnetic field and $\rho$ is the mass density), here representing counter-propagating Alfv\'enic fluctuations, are non-zero. Early simulations \citep{shebalin83} showed MHD turbulence to be anisotropic (and more so at smaller scales), as well as displaying Alfv\'en wave like behaviour, which was explained in terms of nonlinear three-wave interactions. In weak turbulence, where the nonlinear time is slow compared to the Alfv\'en wave time, the modes are usually assumed (and seen in simulations \citep{meyrand16}) to have approximately Alfv\'enic dispersion $\omega=\pm k_\|\vA$, where $k_\|$ is the wavenumber parallel to the mean magnetic field and $\vA$ is the Alfv\'en speed. Under three-wave frequency matching ($\omega_1+\omega_2=\omega_3$) and wavevector matching ($\mathbf{k}_1+\mathbf{k}_2=\mathbf{k}_3$), it can be shown that if all waves maintain this dispersion relation, counter-propagating waves only interact if one of them has $k_\|=0$ (and therefore $\omega=0$) and the $k_\|$ of the other remains unchanged \citep{shebalin83}. There was some debate about whether these interactions would be realisable or whether four-wave interactions would be necessary \citep{sridhar94,goldreich95,montgomery95,ng96,ng97,goldreich97}, but the three-wave interactions (with one $k_\|=0$ mode) are now an accepted component of weak Alfv\'enic turbulence \citep{galtier00,galtier02,lithwick03,boldyrev09a,schekochihin12} and seen in simulations to play an important role \citep{perez08,boldyrev09a,meyrand15}, even if the precise theory is still under discussion \citep{boldyrev09a,schekochihin12,schekochihin22}. More recent work \citep{howes13a,dorfman25} has extended the calculations, showing that a variety of modes, many not satisfying the Alfv\'en wave dispersion relation, are involved in Alfv\'enic turbulence, particularly for stronger interactions.

It was shown \citep{sridhar94,goldreich95,goldreich97} that as weak turbulence cascades energy to smaller scales it becomes strong, and should reach critical balance in which the linear and nonlinear terms remain comparable at all scales. The predicted energy spectrum of strong turbulence, usually $k_\perp^{-5/3}$ or $k_\perp^{-3/2}$ depending on the treatment of the nonlinear timescale \citep{goldreich95,boldyrev06,chandran15,mallet17}, generally match a wide variety of space and astrophysical observations \citep{armstrong95,zhuravleva14,chen16b}, indicating that turbulence in nature is generally strong, as also confirmed by measurements of the strength parameter in the solar wind \citep{chen16b}. Due to localised energy sources and inhomogeneous background conditions, turbulence in nature is also often imbalanced, i.e., has a significantly greater energy flux in Alfv\'enic fluctuations propagating in one direction compared to the other. Models of imbalanced Alfv\'enic turbulence, in both homogeneous and inhomogeneous background conditions, have been proposed, although also remain under discussion \citep{lithwick03,lithwick07,beresnyak08,chandran08,perez09,podesta10c,chandran15,mallet17,chandran19,chandran25}.

Laboratory experiments offer an additional approach \citep{howes18b}, enabling isolated components of the relevant physics to be studied, under repeatable controlled conditions, with a large number of measurement points. Early experiments, such as the Culham Zeta discharge \citep{robinson68,robinson71} and UCLA Macrotor tokamak \citep{zweben79} measured anisotropic fluctuations which were interpreted as MHD turbulence. More recently, the UCLA Large Plasma Device (LAPD) has used various methods to generate Alfv\'en waves, with early work stuyding their linear properties, verifying dispersion relations and damping rates \citep{gekelman94,gekelman97,leneman99,vincena04,palmer05,kletzing03,thuecks09,kletzing10}. Subsequent work has begun to investigate the properties of their nonlinear interactions. Two co-propagating Alfv\'en waves were shown to generate a nonlinear mode \citep{carter06,brugman07}, with the co-propagating nonlinearity interpreted as being due to the slightly dispersive nature of the waves at $k_\perp\rho_\mathrm{s}\approx0.3\textrm{--}0.5$ and $\omega/\Omega_\textrm{ci}\approx0.6$, where $\rho_\mathrm{s}$ is the sound gyroradius and $\Omega_\mathrm{ci}$ the ion cyclotron frequency. The three-wave interactions relevant to the parametric instability have also been seen \citep{dorfman13,dorfman16}. An experiment with counter-propagating Alfv\'en waves was performed \citep{howes12b,drake13}, aimed at creating the three-wave interaction relevant for weak turbulence, in which one of the primary waves had an effective $k_\|=0$ component \citep{howes13b}. The interaction was seen to transfer energy to a new $k_\perp$ and interpreted as evidence for the weak three-wave interaction. A nonresonant interaction was also seen \citep{drake16}, due to the $k_\|\neq0$ components of both primary waves.

There have also been other experiments displaying turbulent signatures in or close to the MHD regime. Alfv\'enic turbulence was reported at the Madison Symmetric Torus \citep{ren11,thuecks17}, where a strong fluctuation anisotropy was found, the Swarthmore Spheromak Experiment \citep{schaffner14a,schaffner14b,schaffner14c,schaffner15}, where broadband power spectra and intermittency were detected, a table-top laser plasma \citep{chatterjee17}, although the fluctuations here were at the ion gyroscale, and the Pegasus Toroidal Experiment \citep{richner22}, which showed a broadband magnetic spectrum with a spectral break. It has remained to be shown, however, what nonlinear processes are responsible and whether an energy cascade is taking place.

In this paper, we report LAPD experiments that reveal the mechanisms involved in Alfv\'enic turbulence by (1) studying the counter-propagating nonlinear interactions relevant to strong turbulence, (2) demonstrating a new co-propagating nonlinearity relevant to imbalanced turbulence, and (3) showing the transfer of energy to smaller perpendicular scales, consistent with a local cascade. It has been a longstanding aim of laboratory plasma astrophysics to create fully-developed Alfv\'enic turbulence under controlled conditions, and this work represents a significant step towards this goal.

\section*{Experimental Setup}

\begin{figure*}
\includegraphics[width=1.8\columnwidth]{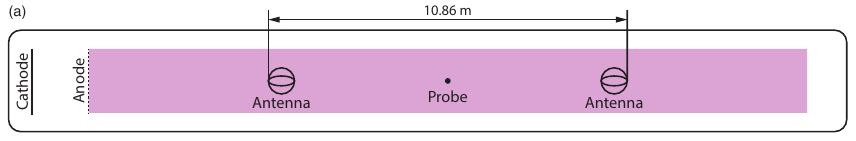}\vspace{0.2cm}
\includegraphics[width=\columnwidth]{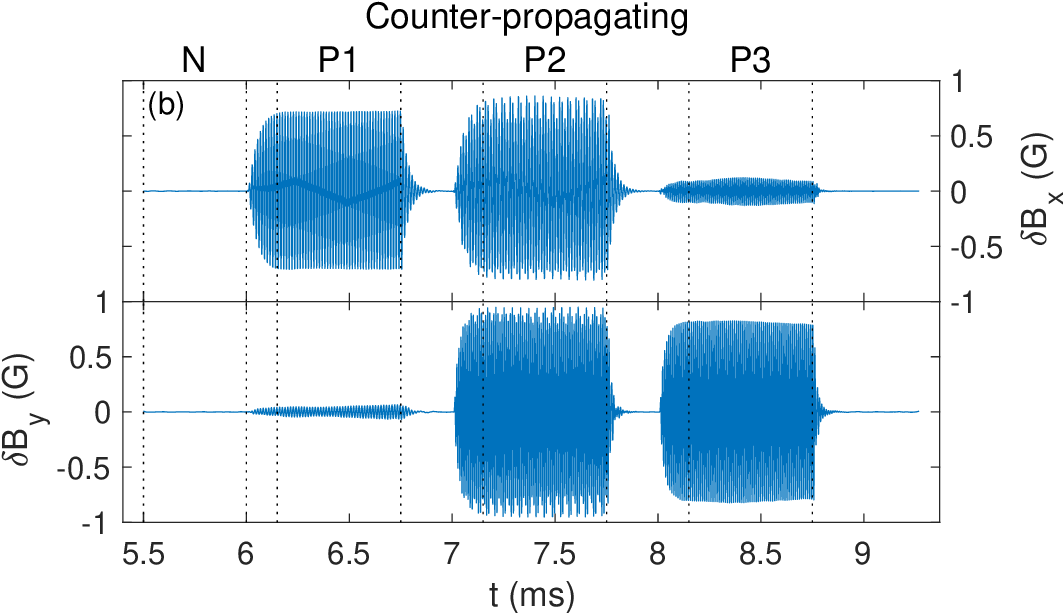}\vspace{0.2cm}
\includegraphics[width=\columnwidth]{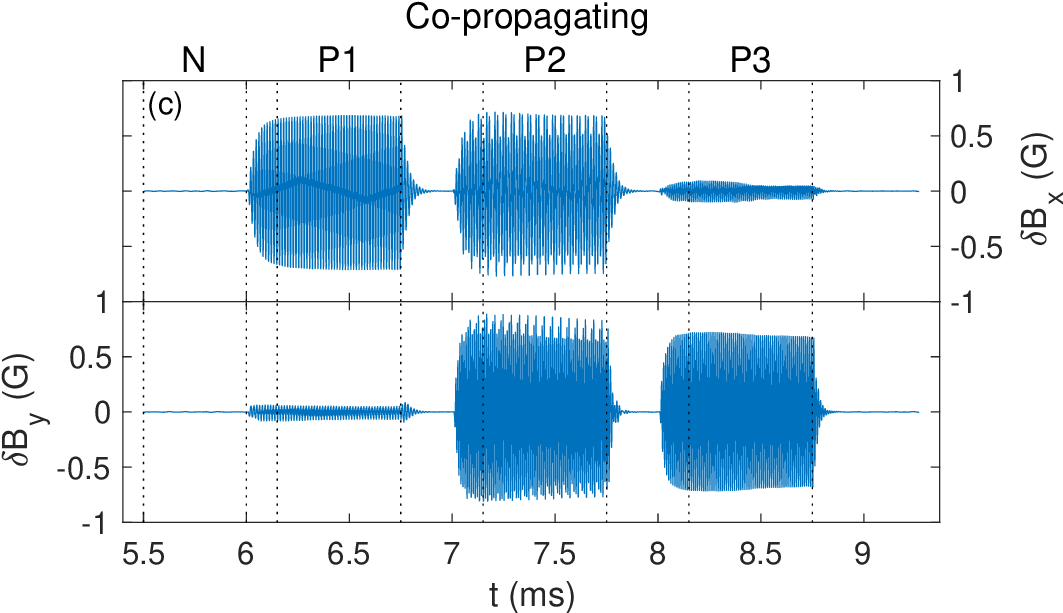}
\includegraphics[width=\columnwidth]{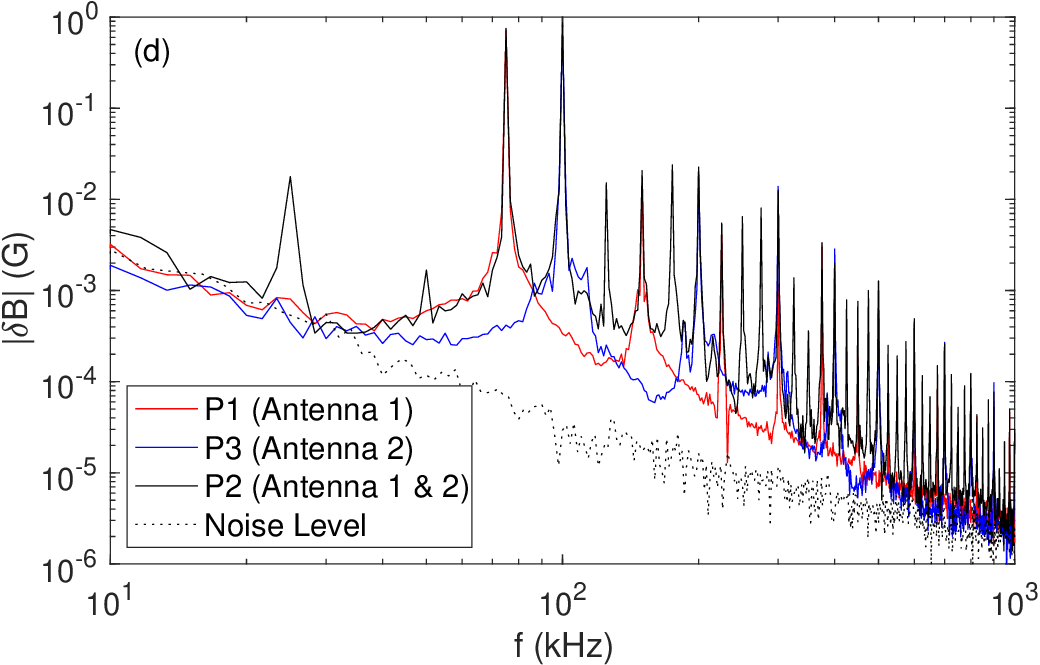}
\includegraphics[width=\columnwidth]{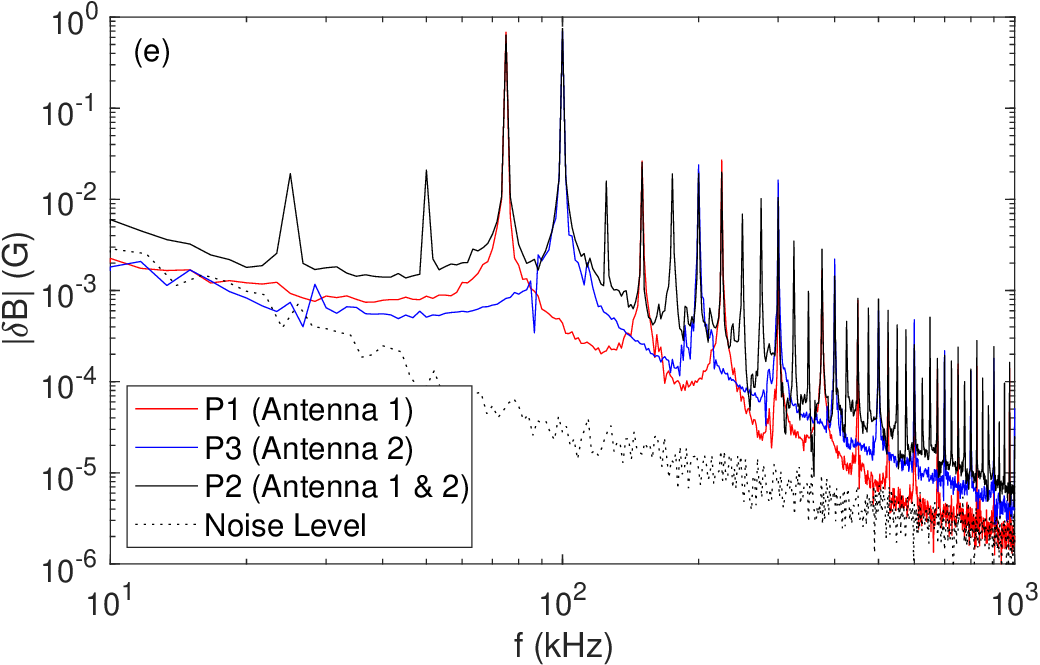}
\caption{\textbf{Experimental setup and magnetic field fluctuation properties during the experiment.} (a) Experimental setup showing the plasma column with antennas and probe. (b) Time series of the fluctuating perpendicular magnetic field components $\delta B_x$ and $\delta B_y$ during the counter-propagating interaction and (c) during the co-propagating interaction. The dotted lines mark the three wave periods: P1 (Antenna 1 only), P2 (Antennas 1\&2), P3 (Antenna 2 only), and the period used for the noise level determination (N). (d) Amplitude spectrum of magnetic fluctuations during the counter-propagating interaction and (e) during the co-propagating interaction in the three wave periods P1, P2, P3. During P2 there is a set of nonlinear modes which are not present when the antennas are on individually (P1\&P3).}
\label{fig:timeseriesandspectra}
\end{figure*}

The experiment was carried out at the Large Plasma Device (LAPD) \citep{gekelman16} in which a cylindrical column of Hydrogen plasma, 60\,cm in diameter and 18\,m long, was made once per second, lasting $\sim$10\,ms each shot. The plasma was created by applying a discharge voltage between a barium oxide (BaO) cathode \citep{leneman06} and a wire mesh anode, 50\,cm form the cathode, at one end of the machine. A constant  background magnetic field $B_0$ was applied along the axis of the device ($z$-direction). The experimental setup is shown in Figure \ref{fig:timeseriesandspectra}(a) and the resulting plasma conditions are given in Table \ref{tab:parameters}.

\begin{table}
\caption{\label{tab:parameters}Experimental parameters (*wave amplitudes are given at the mid-point between antennas and nonlinearity parameters are based on the range of amplitudes along the device)}
\begin{ruledtabular}
\begin{tabular}{ccc}
Parameter & Symbol & Value \\
\hline
mass number & $A$ & 1 (Hydrogen) \\
charge number & $Z$ & 1 \\
magnetic field strength & $B_0$ & 400\,G \\
number density & $n=n_\mathrm{i}=n_\mathrm{e}$ & $8\times10^{11}$\,cm$^{-3}$ \\
electron temperature & $T_\mathrm{e}$ & $\approx$\,4\,eV \\
ion temperature & $T_\mathrm{i}$ & $\approx$\,1\,eV \\
wave 1 frequency & $f_1$ & 75\,kHz \\
wave 2 frequency & $f_2$ & 100\,kHz \\
wave 1 perpendicular wavenumber & $k_{\perp1}$ & 46\,rad\,m$^{-1}$ \\
wave 2 perpendicular wavenumber & $k_{\perp2}$ & 48\,rad\,m$^{-1}$ \\
wave 1 parallel wavenumber & $k_{\|1}$ & 0.47\,rad\,m$^{-1}$ \\
wave 2 parallel wavenumber & $k_{\|2}$ & 0.64\,rad\,m$^{-1}$ \\
wave 1 amplitude* & $\delta B_1$ & 0.75\,G \\
wave 2 amplitude* & $\delta B_2$ & 0.87\,G \\
\hline
wave 1 normalised frequency&$\omega_1/\Omega_\textrm{ci}$ & 0.12 \\
wave 2 normalised frequency&$\omega_2/\Omega_\textrm{ci}$ & 0.16 \\
normalised $k_\perp$&$k_{\perp1,2}\rho_\mathrm{i}$ & 0.17 \\
normalised $k_\perp$&$k_{\perp1,2}\rho_\mathrm{s}$ & 0.24 \\
normalised $k_\perp$&$k_{\perp1,2}d_\mathrm{i}$ & 12 \\
normalised $k_\perp$&$k_{\perp1,2}d_\mathrm{e}$ & 0.28 \\
wave 1 nonlinearity parameter*&$\chi_1$ & 0.42 -- 0.68 \\
wave 2 nonlinearity parameter*&$\chi_2$ & 0.15 -- 0.27 \\
ion beta&$\beta_\mathrm{i}$&$2\times10^{-4}$\\
electron beta&$\beta_\mathrm{e}$&$8\times10^{-4}$ \\
total beta&$\beta$&$1\times10^{-3}$ \\
\end{tabular}
\end{ruledtabular}
\end{table}

Two rotating magnetic field (RMF) antennas \citep{gigliotti09} were placed in the plasma 10.86\,m apart to generate Alfv\'en waves propagating along the device. Each antenna has two orthogonal loops, $\approx9\,\textrm{cm}$ in diameter, that can be driven independently, allowing it to produce a wave polarised in the $x$-direction, one polarised in the $y$-direction, or both at the same time. Two configurations were investigated: a counter-propagating interaction, in which one wave was launched from each antenna to interact between the antennas, and a co-propagating interaction, in which two waves were launched from the same antenna to interact as they propagate away. In both cases, the waves were perpendicularly polarised and at frequencies $f_1=75\,\mathrm{kHz}$ and $f_2=100\,\mathrm{kHz}$. Each shot consisted of three time periods: wave 1 only (6.00--6.75\,ms), both waves (7.00--7.75\,ms) and wave 2 only (8.00--8.75\,ms), where time is measured from the start of the discharge.

\begin{figure*}
\includegraphics[width=\textwidth]{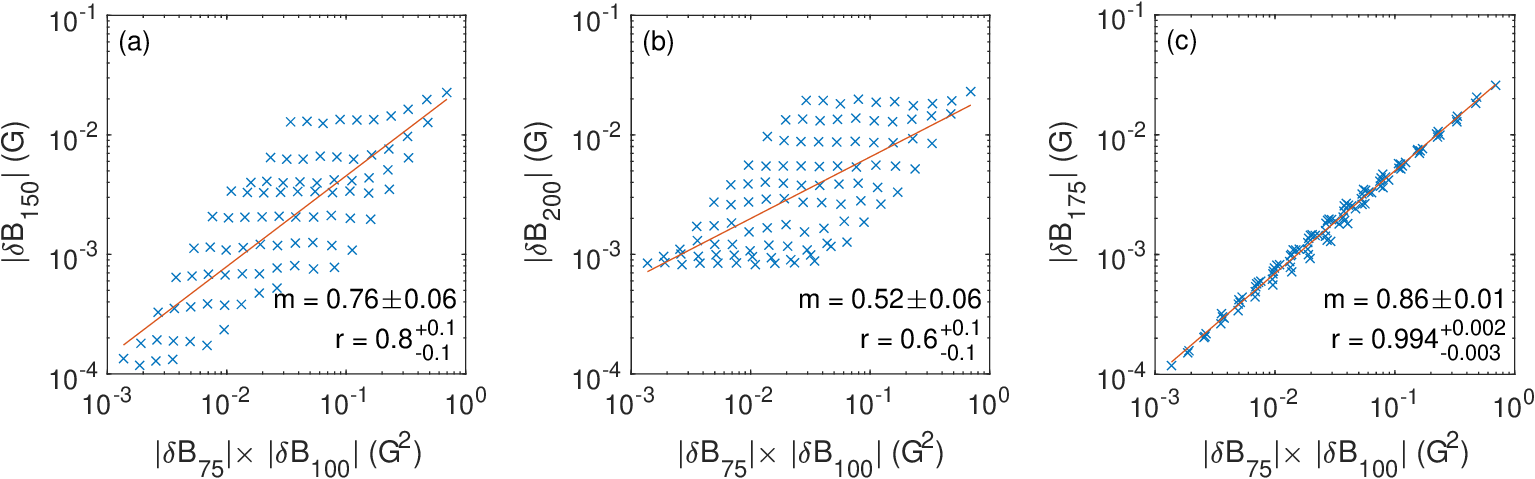}
\includegraphics[width=\textwidth]{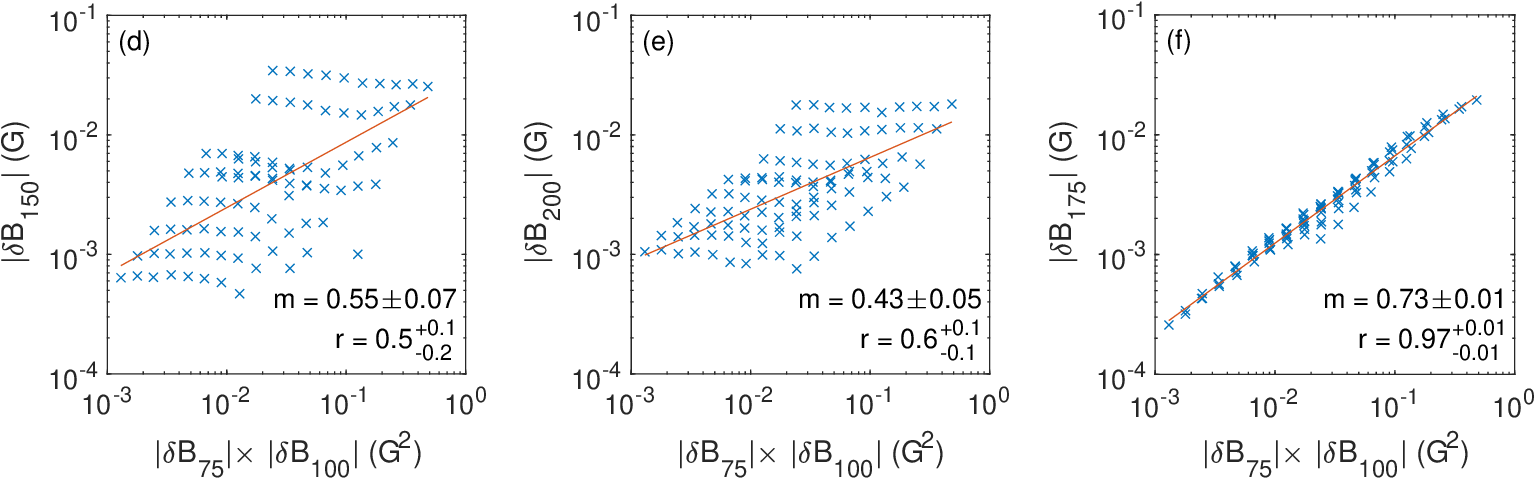}
\caption{\textbf{Dependence of several mode amplitudes on the product of the primary wave amplitudes.} (a-c) Amplitudes of the 150\,kHz, 200\,kHz, and 175\,kHz modes in the counter-propagating interaction during period P2 as a function of the product of the primary wave amplitudes, 75\,kHz and 100\,kHz, with fitted power laws marked as solid red lines, their the power law indices, $m$, with the standard error indicated, and the correlation coefficients, $r$, and their 95\% confidence intervals indicated. (d-f) Same for the co-propagating interaction. In both cases, the antenna harmonics (150\,kHz and 200\,kHz) form diamond patterns, expected if they depend only on their single fundamental mode, whereas the nonlinear mode at the sum frequency 175 kHz depends on the product of both primary waves, as expected from a quadratic nonlinearity.}
\label{fig:amplitudescan}
\end{figure*}

The magnetic field fluctuations were measured with a magnetic induction probe (``B-dot probe'') \citep{bose19b} placed in between the antennas. Multiple shots were taken with the probe at varying locations to map the structure of the interactions and the measured amplitudes and wavenumbers of the primary waves are given in Table \ref{tab:parameters}. The $k_\perp$ values were determined from Bessel function fits \citep{brugman07} to the wave patterns in the $x$-$y$ plane when the single waves were present and the $k_\|$ values calculated from the Alfv\'en wave dispersion relation. The normalised frequency and $k_\perp$ values are small ($\ll1$), with the exception of $k_\perp d_\mathrm{i}$, where $d_\mathrm{i}$ is the ion inertial length, meaning that the linear Alfv\'en waves should satisfy $\omega\approx\pm k_\|v_\mathrm{A}$. To measure interactions relevant to strong MHD turbulence, the nonlinearity parameters $\chi_{1,2}=(\delta z_{2,1}k_{\perp1,2})/(v_\mathrm{A}k_{\|1,2})$, where $\delta z_{2,1}$ are the Elsasser amplitudes \citep{elsasser50}, were made as large as possible (while keeping the above non-MHD terms small), giving values of $\chi_{1,2}$ approaching unity. The transition from weak to strong turbulence has been seen in simulations at $\chi\approx1/3$ \citep{meyrand16}, although given that our experiments are not driven with a $k_\|=0$ mode, they naturally probe interactions outside classical weak turbulence theory. A time series of the perpendicular magnetic field components during both interactions, taken at the mid-point between antennas and the centre of perpendicular plane $(0,0,0)$, and band-pass filtered in the range 10\,kHz--20,000\,kHz, is shown in Figure \ref{fig:timeseriesandspectra}(b,c).

\section*{Nonlinear Wave Interactions}

Figure \ref{fig:timeseriesandspectra}(d,e) shows the amplitude spectra of magnetic fluctuations at $(0,0,0)$ during each of the time periods for both counter- and co-propagating cases. The spectra were determined by averaging the time series over 100 shots (to reduce noise while maintaining the wave signals, since the primary waves are highly coherent between shots) then calculating the fast Fourier-transform (FFT) of the averaged time series in each period. When only wave 1 is launched (during P1) clear peaks at its frequency and harmonics can be seen, and similarly for wave 2 (P3). However, when both waves are launched (P2) additional peaks can be seen, as a result of the nonlinear interaction. Since neither antenna-driven wave has $k_\|=0$ and the additional modes are at different frequencies, this is clear evidence for nonlinear interactions relevant to strong turbulence. The modes occur at multiples of the difference frequency, 25\,kHz, as expected from frequency matching conditions, and extend to frequencies $\gtrsim f_\mathrm{ci}$ where two-fluid effects become important.

To investigate the nature of the nonlinearities, the amplitudes of the two primary waves were varied in a logarithmically spaced $10\times10$ grid to give 100 different amplitude settings. The amplitudes of three modes are shown in Figure \ref{fig:amplitudescan} as a function of the product of the primary wave amplitudes, for both the counter- and co-propagating cases, for values a factor of 2 above the noise level. It can be seen that the antenna harmonics, 150\,kHz and 200\,kHz, behave differently to the sum frequency mode, 175\,kHz. The sum mode depends only on the product of the primary waves, as expected for a quadratic nonlinearity, whereas the antenna harmonics form a diamond pattern, expected if the mode depends on one primary wave only, i.e., the fundamental it is a harmonic of. The power law indices, $m$, for the sum modes are, however, slightly shallower than one, indicating a reduced efficiency of the nonlinearity. In the co-propagating case, $m$ is slightly lower than in the counter-propagating case, and the correlation, $r$, is also slightly lower, perhaps suggesting contributions from other interaction paths beyond the quadratic nonlinearity.

\section*{Co-Propagating Interaction Mechanism}

\begin{figure*}
\includegraphics[scale=0.44]{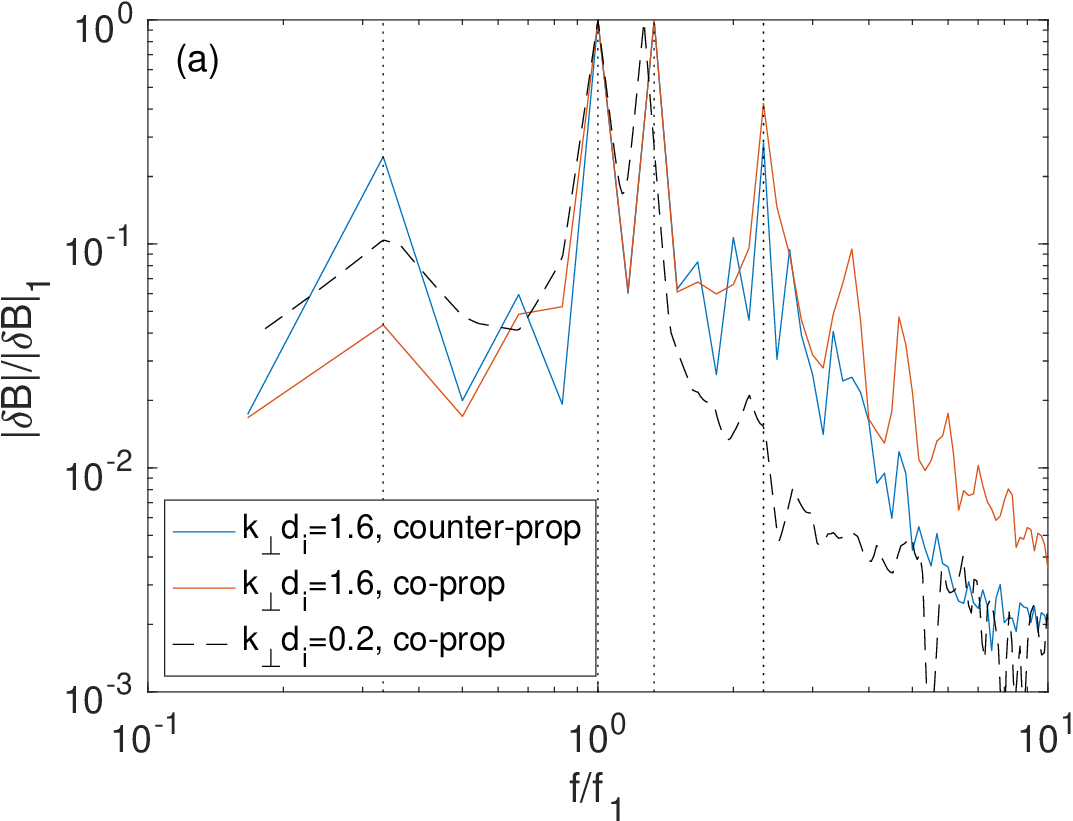}\hspace{1cm}
\includegraphics[scale=0.44]{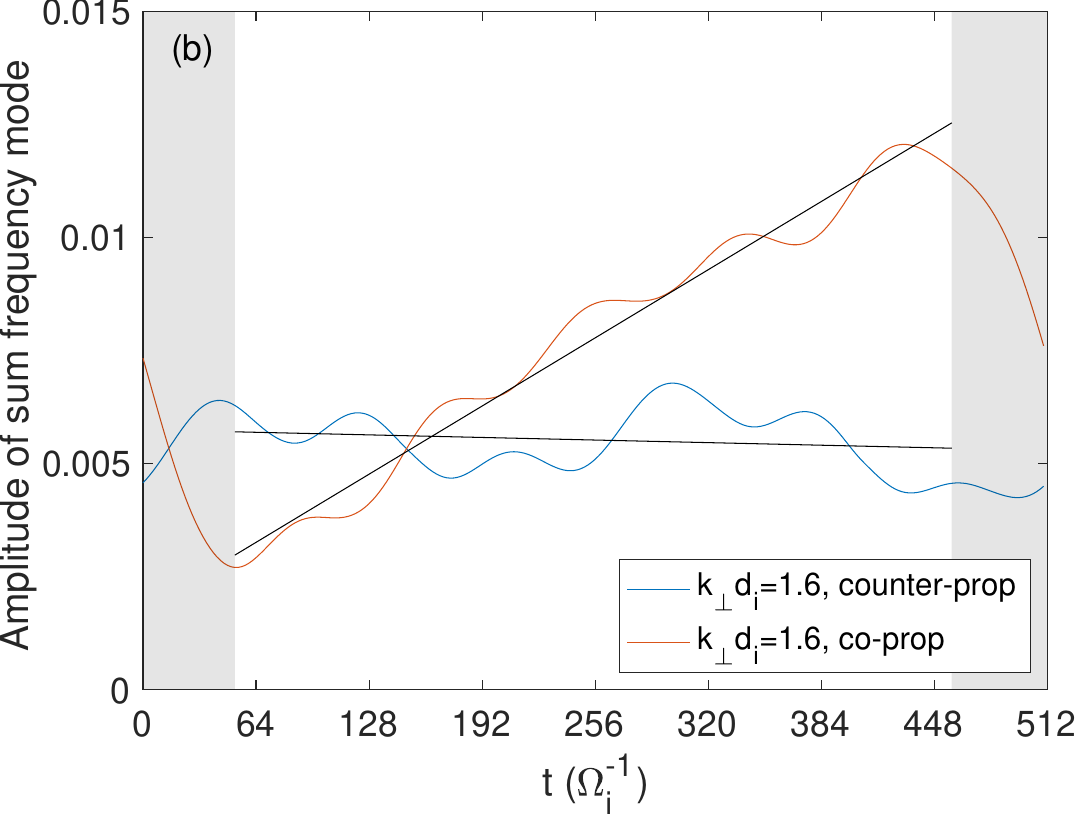}
\caption{\textbf{Spectra and time dependence of magnetic fluctuations in wave interaction simulation.} (a) Amplitude spectrum of magnetic field fluctuations in the 3D hybrid simulations, normalised to the amplitude and frequency of the lower frequency primary wave ($f_1$); the primary wave frequencies and sum/difference frequencies are marked with dotted lines. There is a strong peak at the sum frequency in both counter-propagating and co-propagating interactions at $k_\perp d_\mathrm{i}=1.6$, but not in the co-propagating case at $k_\perp d_\mathrm{i}=0.2$. (b) Amplitude of the sum frequency mode, $f_1+f_2$, in the 3D hybrid simulations at $k_\perp d_\mathrm{i}=1.6$ as a function of time $t$, with straight line fits. The shaded regions mark the wavelet cone of influence \citep{torrence98}. Secular growth is seen in the co-propagating case but not in the counter-propagating case.}
\label{fig:simulation}
\end{figure*}

A spectrum of nonlinear interactions is expected in the counter-propagating case, since the nonlinear MHD term is non-zero, but might not be expected in the co-propagating case, since even though Alfv\'en waves develop density fluctuations at $k_\perp d_\mathrm{i}\gtrsim1$ \citep{hollweg99}, there is no first-order co-propagating nonlinearity when all other non-MHD terms are small \citep{chen14b,boldyrev15,mallet23}. To explain the observations, we turn to a recent theoretical investigation into the effects of finite $k_\perp d_\mathrm{i}$ on nonlinear interactions \citep{mallet23}. Starting from Hall MHD, applying the reduced ordering (keeping $k_\perp d_\mathrm{i}\sim1$), and keeping the two lowest orders, the nonlinear equations become:
\begin{align}
&\ddt\nap^2\zeta_2^\pm\mp\vA\ddz\nap^2\zeta_2^\pm \nonumber\\
&=-\left[\brak{\zeta_2^-}{\nap^2\zeta_1^+}+\brak{\zeta_2^+}{\nap^2\zeta_1^-}+\brak{\zeta_1^-}{\nap^2\zeta_2^+}\right.\nonumber\\
&+\left. \brak{\zeta_1^+} {\nap^2\zeta_2^-}\mp\nap^2\left(\brak{\zeta_1^+}{\zeta_2^-}+\brak{\zeta_2^+}{\zeta_1^-}\right)\right]/2\nonumber\\
&+\left[\nap\nap^2\phi_1\cdot \nap \nap^{-2}\brak{\psi_1}{\nap^2\psi_1}\right.\nonumber\\
&-\left.\nap^2\psi_1\brak{\psi_1}{\nap^2\phi_1}+\brak{\nap^2\phi_1}{b_{z1}\vA}\right]/\Omega_\textrm{ci}\nonumber\\
&\pm\left[\nap^2\brak{\psi_1}{b_{z1}\vA}-\nap^2\left(\nap\psi_1\cdot\nap\nap^{-2}\brak{\psi_1}{\nap^2\psi_1}\right)\right.\nonumber\\
&\left.+\vA\ddz\nap\cdot\left((\hvz\times\nap\phi_1) \cdot \nap (\hvz\times\nap\phi_1)\right)\right]/\Omega_\textrm{ci},
\label{eq:els2ord}
\end{align}
where $\zeta^\pm$ are the Elsasser potentials, $\phi$ is the flux function, $\psi$ the stream function, $\Omega_\textrm{ci}$ the ion cyclotron frequency and numerical subscripts represent the fluctuation order. It can be seen that the terms containing $\zeta^\pm_{1,2}$ are the usual reduced MHD \citep{strauss76,schekochihin09}, and the other terms (both quadratic and cubic nonlinearities) describe a nonlinear drive for $\zeta^\pm_2$. Crucially, even though the linear Alfv\'en waves of this system satisfy $\omega=\pm k_\|v_\mathrm{A}$, the form of these additional terms means that a single Alf\'ven wave, or a set of co-propagating Alfv\'enic fluctuations, are no longer exact nonlinear solutions, so interactions between co-propagating waves can occur. Given that the experimental regime (all non-MHD terms small except $k_\perp d_\mathrm{i}$, low $\beta$, $k_\|/k_\perp\sim\delta B/B_0\ll1$) matches that of the above model, this provides a possible explanation for the observed co-propagating interaction. 

\begin{figure*}
\includegraphics[width=\textwidth]{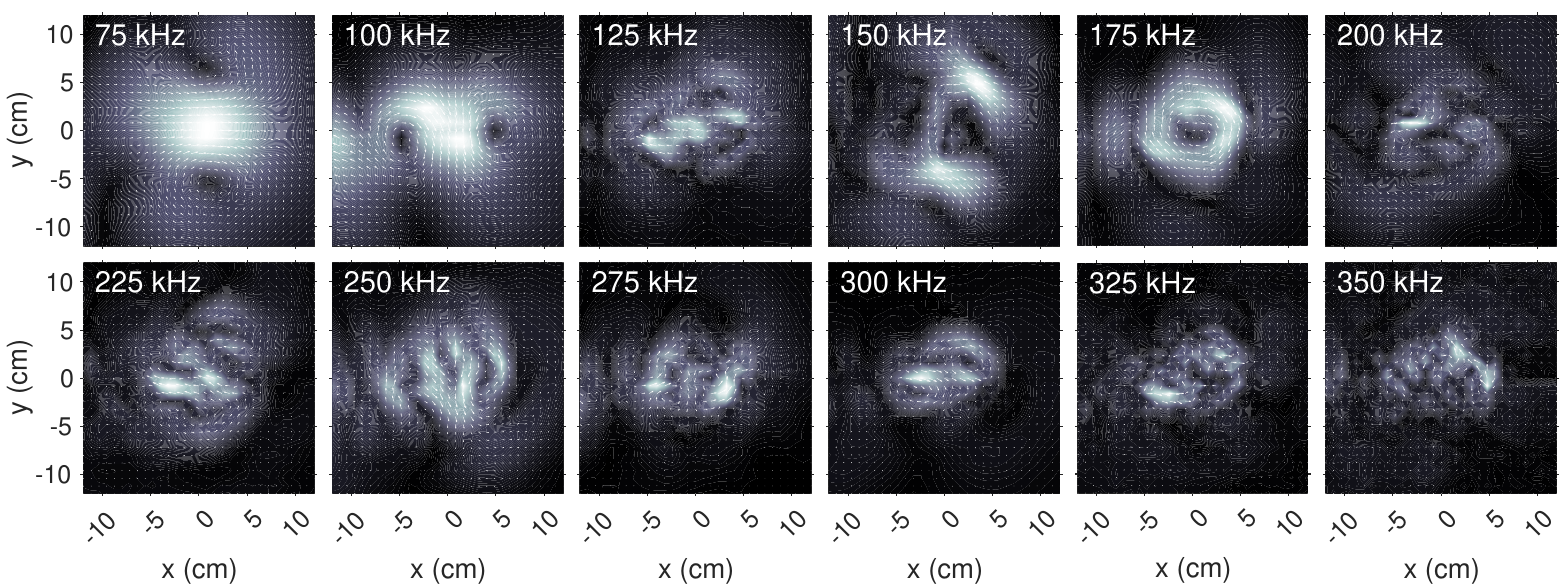}
\includegraphics[width=\textwidth]{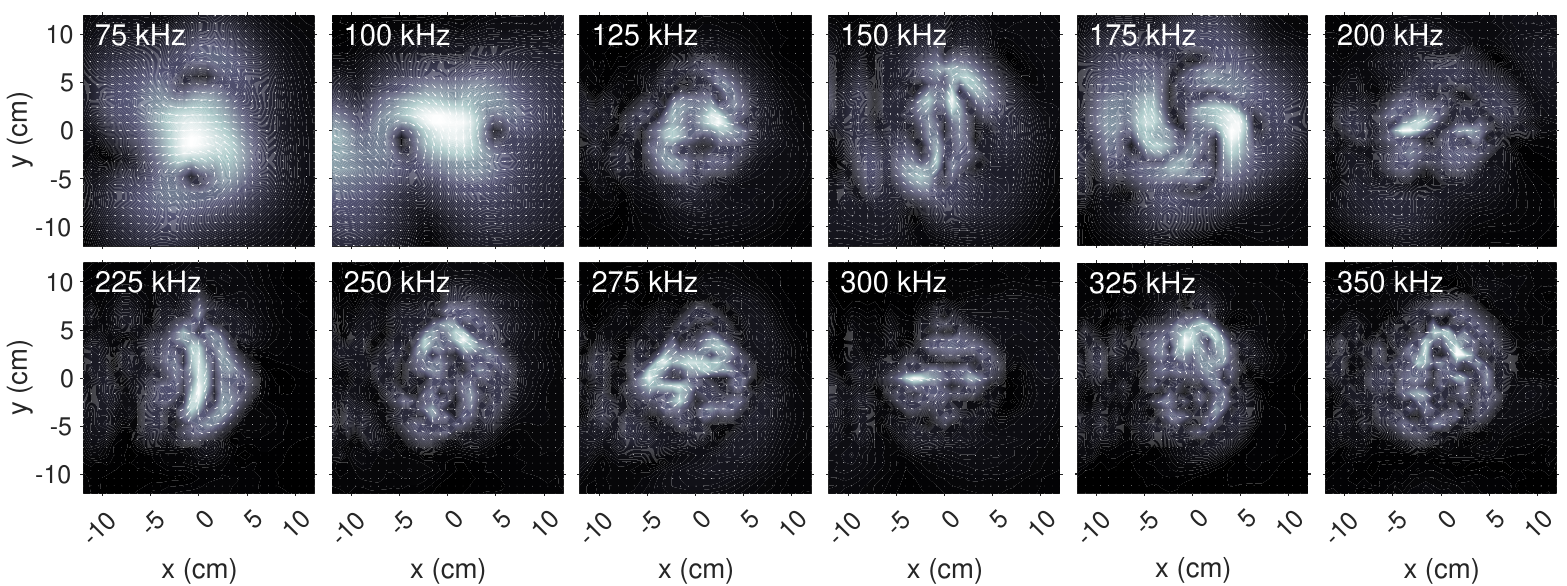}
\caption{\textbf{Wave patterns for a selection of modes in the perpendicular plane.} Amplitude of fluctuations in the perpendicular plane $(x,y)$ during period P2 filtered at the different mode frequencies for the counter-propagating (upper two rows) and co-propagating (lower two rows) interaction. The images are averages over all periods of each mode, and are taken at the phase at which the amplitude is maximum. Arrows (at the measured probe locations) indicate the magnitude and direction of $\delta\mathbf{B}_\perp$, and the color scale, from dark to light, indicates $|\delta\mathbf{B}_\perp|$, and is normalised to the maximum in each image. A general trend towards smaller scale structure for higher frequency modes can be seen.}
\label{fig:wavepatterns}
\end{figure*}

To test this possibility, we further examined the predictions of the model. Considering a 3-wave interaction with $\chi\ll1$, it can be shown \citep{mallet23} that two co-propagating $\zeta^+$ waves of similar amplitudes and scales produce a nonresonant $\zeta^-$ mode with amplitude given by:
\begin{equation}
\frac{z_2^-}{v_\mathrm{A}}\sim k_\perp d_\mathrm{i}\sin^2\alpha\left(\frac{z_1^+}{v_\mathrm{A}}\right)^2,
\label{eq:nonresonantamplitude}
\end {equation}
and a resonant $\zeta^+$ mode with amplitude given by:
\begin{equation}
\frac{z_2^+}{v_\mathrm{A}}\sim k_\perp d_\mathrm{i}\sin^2\alpha\left(\frac{z_1^+}{v_\mathrm{A}}\right)^2k_\|v_\mathrm{A}t.
\label{eq:resonantamplitude}
\end {equation}
Using the experimental parameters: $z_1^+=4.0\times10^3\,\mathrm{m}\,\mathrm{s}^{-1}$ (the geometric mean of the primary wave amplitudes), $v_\mathrm{A}=9.8\times10^5\,\mathrm{m}\,\mathrm{s}^{-1}$, $k_\perp d_\mathrm{i}=12$, $\sin^2\alpha=1$, $k_\|=0.55\,\mathrm{rad}\,\mathrm{m}^{-1}$ (the average parallel wavenumber of the primary waves), $t=5.6$\,\textmu s (the time for the waves to travel from the antenna to the probe at $(0,0,0)$), the predicted amplitudes of these modes are $z_2^-/z_1^+=0.05$ and $z_2^+/z_1^+=0.15$, i.e., the resonant mode should be higher amplitude but still satisfies $z_2^+/z_1^+\ll1$ so the calculation remains valid. This value can now be compared to the experimentally measured value of $z_2/z_1$. The geometric mean of the primary waves at their peak location, $(x=0,y=0)$, is 0.78\,G and the amplitude of the sum mode 175 kHz at its peak location, $(x=4.8,y=0)$\,cm, is 0.090\,G, giving $z_2/z_1=0.12$, close to the predicted value of 0.15 for the resonant mode. Given that the theoretical prediction and experimental value are only expected to match to order unity, this is good evidence that the co-propagating interaction is due to these higher-order finite $k_\perp d_\mathrm{i}$ nonlinearities.

\section*{Numerical Simulations}

To further test the origin of the nonlinear interactions, we ran 3D hybrid particle-in-cell simulations, with kinetic ions and a massless electron fluid, using the CAMELIA code \citep{franci18}. A $256^3$ simulation box was used, with grid spacing $\Delta x,\Delta y=0.016 d_\mathrm{i}$, $\Delta z=2d_\mathrm{i}$, a time step $\Delta t=0.01\Omega_\textrm{ci}^{-1}$, and 512 particles per cell (ppc). The simulation was initialised with counter- and co-propagating plane Alfv\'en waves. The plasma and wave conditions were chosen to be in a similar regime to the experiment: $\beta_\mathrm{i,e}=10^{-3}$, $\delta B_{1,2}/B_0=0.02$, $\omega_1/\Omega_\textrm{ci}=0.074$, $\omega_2/\Omega_\textrm{ci}=0.098$, $k_{\perp 1,2}\rho_\mathrm{i}=0.050$, $k_{\perp 1,2}\rho_\mathrm{s}=0.035$, $k_{\perp 1,2}d_\mathrm{i}=1.57$, $(k_\perp/k_\|)_1=21.3$, $(k_\perp/k_\|)_2=16$, $\chi_1=0.85$, $\chi_2=0.64$. Spectra of the magnetic fluctuations, taken and averaged over several points in the simulation, are shown in Figure \ref{fig:simulation}(a), where it can be seen that in both of these $k_\perp d_\mathrm{i}\sim1$ cases a strong peak is produced at the sum frequency $f_1+f_2$, as well as a broader spectrum at higher frequencies. This is consistent with the experiment in which both counter- and co-propagating waves interact nonlinearly. 

The simulation allowed us to test the co-propagating interaction model by reducing $k_\perp d_\mathrm{i}$, which was not possible in the experiment. A further simulation was run with $k_{\perp 1,2}d_\mathrm{i}=0.20$. For this simulation, the box size was $128^3$, $\Delta x,\Delta y=0.25\,d_\mathrm{i}$, $\Delta z=16d_\mathrm{i}$, $\Delta t=0.05\Omega_\textrm{ci}^{-1}$, $\textrm{ppc}=4096$, $\beta_\mathrm{i,e}=0.1$, $\omega_1/\Omega_\textrm{ci}=0.018$, $\omega_2/\Omega_\textrm{ci}=0.024$, $k_{\perp 1,2}\rho_\mathrm{i}=0.062$, $k_{\perp 1,2}\rho_\mathrm{s}=0.044$, $(k_\perp/k_\|)_1=10.7$, $(k_\perp/k_\|)_2=8$, which give the same strength parameters $\chi_1=0.85$, $\chi_2=0.64$. For this $k_\perp d_\mathrm{i}\ll1$ co-propagating interaction, it can be seen from Figure \ref{fig:simulation}(a) that the strong peak at the sum frequency is reduced by an order of magnitude, and the spectrum at higher frequencies is much lower. This shows that the nonlinear interaction is controlled by $k_\perp d_\mathrm{i}$ as predicted by the model \citep{mallet23}. 

Another key prediction of the model is that the resonant mode amplitude grows in time according to Equation (\ref{eq:resonantamplitude}). To test this in the simulations, a wavelet transform \citep{torrence98} was performed and the amplitude of the sum frequency mode shown as a function of time in Figure \ref{fig:simulation}(b). In the counter-propagating case, the mode amplitude is constant, as expected for the standard MHD nonlinearity, however, in the co-propagating case the mode grows in time. The fit gradient for the co-propagating case is $2.4\times10^{-5}\pm2.5\times10^{-7}$, and the theoretical value (from Equation (\ref{eq:resonantamplitude})) is $1.7\times10^{-5}$; given that the prediction should be correct to order unity, this is a good match, supporting this explanation for the co-propagating interaction.

\section*{Perpendicular Energy Transfer}

One of the aims of the experiment was to drive sufficiently large $\chi$ to create a measurable and continuous transfer of energy to higher $k_\perp$. To measure this, a B-dot probe was scanned over the $x$-$y$ plane in a $41\times41$ grid covering positions up to $x,y=\pm16$\,cm, with 15 shots taken in each location. Images of the wave patterns are shown in Figure \ref{fig:wavepatterns}, created by first averaging the time series over the 15 shots during P2, band-pass filtering at each frequency $\pm10\,$kHz, averaging over all periods at each frequency, and plotting $\delta\mathbf{B}_\perp$ for the phase at which the maximum amplitude occurs.

In both counter- and co-propagating cases, the primary waves (75\,kHz and 100\,kHz) have a simple large-scale structure with two current channels, as expected from the antennas \citep{gigliotti09}, and as the frequency increases there is a trend towards smaller scale structure. Since higher frequencies (as a general trend) result from a larger number of interactions, this shows that as the number of interactions increases, smaller scale perpendicular structure is generated. This is evidence for energy transfer consistent with a perpendicular cascade, i.e., local energy transfer to progressively smaller perpendicular scales.

\begin{figure}
\includegraphics[width=\columnwidth]{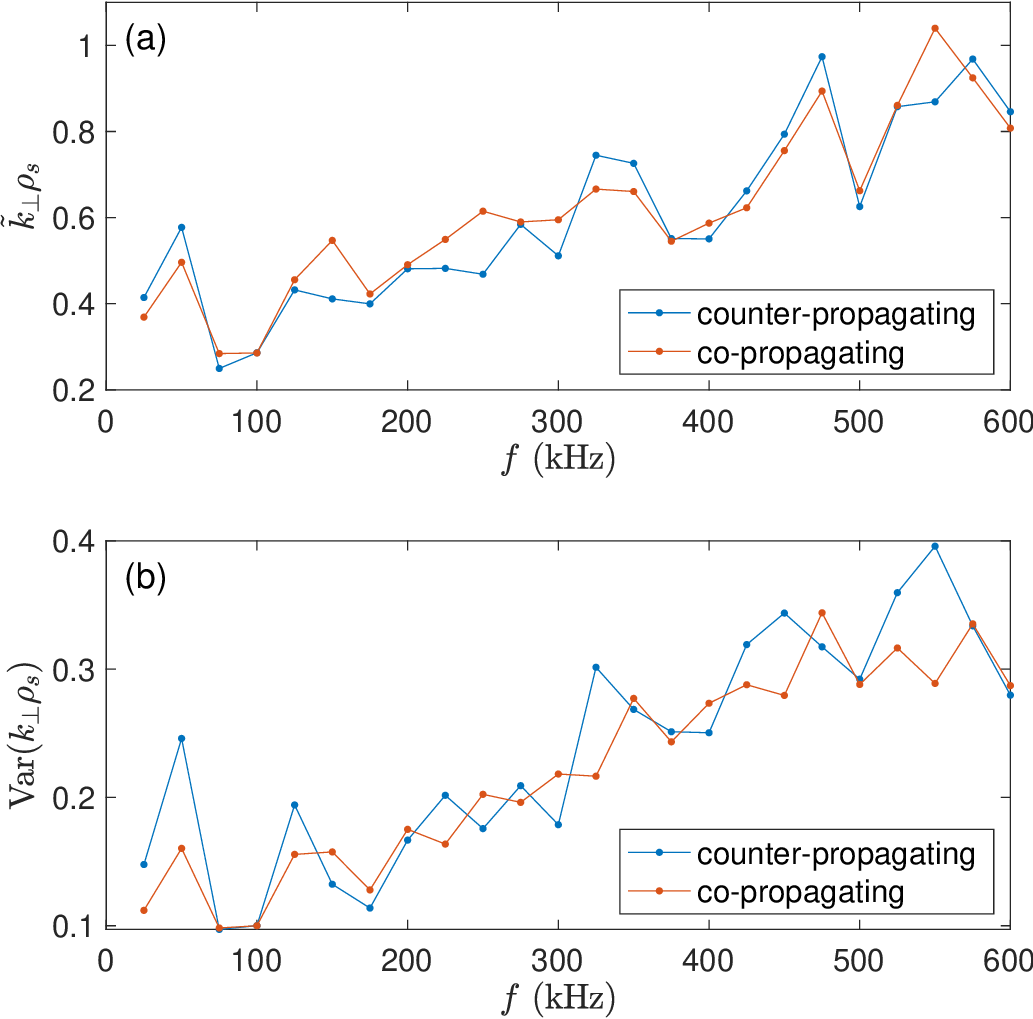}
\caption{\textbf{Typical scale and spread of energy as a function of mode frequency.} (a) Typical fluctuation scale $\tilde{k}_\perp\rho_\mathrm{s}$ (Equation~ \ref{eq:avk}) and (b) spread of energy over $k_\perp$ (Equation~ \ref{eq:vark}), as a function of mode frequency $f$ for both counter-propagating and co-propagating interactions. In both cases, there is a tendency for the higher frequency modes to have larger $\tilde{k}_\perp\rho_\mathrm{s}$ values and their energy spread over a wider range of $k_\perp$, consistent with the development of a perpendicular cascade.}
\label{fig:kperpstructure}
\end{figure}

To quantify this, a 3D FFT was calculated over the region $(\pm8,\pm8,0)$\,cm during P2 to give $\delta\mathbf{B}(f,k_x,k_y)$, then for the frequencies $f$ with wave power (multiples of 25\,kHz), the 2D FFT $\delta\mathbf{B}(k_x,k_y)$ was averaged over the azimuthal angle in the $(k_x,k_y$) plane to give $\delta\mathbf{B}(k_\perp)$, and the quantities
\begin{equation}
\tilde{k}_\perp=\frac{\int_0^\infty k_\perp|\delta\mathbf{B}(k_\perp)|dk_\perp}{\int_0^\infty|\delta\mathbf{B}(k_\perp)|dk_\perp}
\label{eq:avk}
\end{equation}
and
\begin{equation}
\mathrm{Var}({k}_\perp)=\frac{\int_0^\infty (k_\perp-\tilde{k}_\perp)^2|\delta\mathbf{B}(k_\perp)|dk_\perp}{\int_0^\infty|\delta\mathbf{B}(k_\perp)|dk_\perp}
\label{eq:vark}
\end{equation}
were calculated, which represent a weighted average $k_\perp$ at which the energy lies, and the spread of energy over $k_\perp$ (note that here $k_\perp^2=k_x^2+k_y^2$ which is different to that used in the Bessel function fit values in Table \ref{tab:parameters}). These are shown, normalised to $\rho_\mathrm{s}$, in Figure \ref{fig:kperpstructure} up to 600\,kHz, beyond which the amplitudes begin to be affected by instrumental noise. It can be seen that in both counter- and co-propagating cases, the primary waves (75\,kHz and 100\,kHz) have the lowest $\tilde{k}_\perp$ and $\mathrm{Var}({k}_\perp)$ values, the difference and sum modes have the next lowest values, then there is a gradual increase by a factor of $\approx3$ in both $\tilde{k}_\perp$ and $\mathrm{Var}({k}_\perp)$ over this frequency range. This is consistent with energy being transferred to smaller perpendicular scales as the waves interact and the energy being spread over a broader $k_\perp$ spectrum. While this is not yet fully developed turbulence, and the higher frequency modes are relatively low amplitude (Figure \ref{fig:timeseriesandspectra}), this measurement of energy transfer to progressively small perpendicular scales represents a significant step towards the generation of controlled Alfv\'enic turbulence in the laboratory.

\section*{Discussion}

In this experiment, we have shown several important features of nonlinear interactions relevant to Alfv\'enic turbulence. We have demonstrated that the counter-propagating interactions relevant to strong MHD turbulence occur and generate a wide frequency spectrum of nonlinear modes, some of which are expected to be nonresonant modes, i.e., not following the Alfv\'en wave dispersion \citep{howes13a,dorfman25}, and with the expected quadratic scaling for the initial sum mode. We have shown that co-propagating non-dispersive Alfv\'en waves can interact, and generate nonlinear modes in agreement with recently discovered finite $k_\perp d_\mathrm{i}$ nonlinearities arising from higher order terms in Hall MHD \citep{mallet23}. We have also shown that in both cases, energy is progressively transferred to, and spread over, smaller perpendicular scales, consistent with a perpendicular cascade, suggesting that both resonant and nonresonant Alfv\'en modes can play a role in strong turbulence.

While a previous experiment \citep{carter06,brugman07} measured a co-propagating Alfv\'en wave interaction, the physics probed here is different. In the previous experiment, the waves were closer to the kinetic regime, $k_\perp\rho_\mathrm{s}\approx0.3\textrm{--}0.5$ and $\omega/\Omega_\textrm{ci}\approx0.6$, and the interaction was interpreted as due to their dispersive nature, for which the standard kinetic Alfv\'en nonlinearity applies \citep{voitenko98,cho11,voitenko11,voitenko16}. However, here these parameters are smaller and cannot account for the generation of nonlinear modes at a level comparable to the counter-propagating case (Figure \ref{fig:timeseriesandspectra}). Instead, we have shown that their cause is a new $k_\perp d_\mathrm{i}$ nonlinearity, where the waves remain non-dispersive.

There are a number of important implications of this work for space and astrophysical systems. Many of these systems, most notably the solar wind, contain large scale Alfv\'en waves, and we have shown experimentally that the interaction of such waves can transfer energy to smaller scales, i.e., we have confirmed that they can be an important source of the turbulence in these systems. The co-propagating interaction of non-dispersive Alfv\'en waves may be particularly important in regions where the imbalance is large -- it is expected to dominate over the counter-propagating MHD interaction when $\delta z^-/\delta z^+\lesssim k_\perp d_\mathrm{i}\delta z^+/v_\mathrm{A}$ \citep{mallet23}. In space and astrophysical systems with a localised source, such as the near-Sun solar wind \citep{chen20}, this nonlinearity may therefore dominate if $k_\perp d_\mathrm{i}$ is not too small, and allow a cascade to occur, even at very large imbalance. The presence of a helicity barrier \citep{meyrand21,squire22,mcintyre25}, which would increase $\delta B/B$, may also enhance this interaction, and, like the helicity barrier, these nonlinearities may enable high-frequency heating mechanisms, since they can couple fluctuations to higher $k_\|$ \citep{mallet23}. However, it is also possible that this interaction may weaken the barrier, in allowing some portion of the imbalanced energy to cascade through ion scales; this interplay will need to be the subject of future study. Similar finite $k_\perp d_\mathrm{i}$ nonlinearities may also help to explain the strong magnetic field gradients often seen in the solar wind \citep{mallet23}, and may help to explain the break seen in the solar wind turbulence spectrum at $k_\perp d_\mathrm{i}\sim1$ \citep{chen14b}.

Finally, this work represents a significant step towards the generation of controlled fully-developed Alfv\'enic turbulence in the laboratory to provide a complete understanding of it as a fundamental plasma process. We have shown that both counter- and co-propagating Alfv\'en waves can transfer energy to smaller perpendicular scales, that different nonlinearities responsible for energy transfer can be selectively studied, and that experiments, theory, and simulations together can be used to understand the nature of space and astrophysical plasma turbulence.

\begin{acknowledgments}
CHKC and SG were supported by UKRI Future Leaders Fellowship MR/W007657/1. CHKC was also supported by STFC Consolidated Grant ST/X000974/1. CHKC and LF were supported by STFC Consolidated Grant ST/T00018X/1. SD was supported by DOE grant DE-SC0021291 and NSF grant AGS-2401219. SB was supported by the U.S. Department of Energy, Office of Science, Office of Fusion Energy Sciences award number DE-SC0024362. LF was supported by Royal Society University Research Fellowship URF/R1/231710 and Royal Society International Exchanges award IES\textbackslash R1\textbackslash 231251. This experiment was performed at the UCLA Basic Plasma Science Facility supported by DOE and NSF. This project was supported by Royal Society International Exchanges award IES\textbackslash R2\textbackslash 170137. This work used the ARCHER UK National Supercomputing Service (http://www.archer.ac.uk) and the DiRAC-2.5y Director’s Discretionary allocation ``Plasma turbulence in the low electron beta regime'' (PI: Franci); DiRAC is part of the UKRI Digital Research Infrastructure (http://www.dirac.ac.uk). We acknowledge the UCLA BaPSF staff for useful discussions and experimental assistance, and D. Burgess for simulation support.
\end{acknowledgments}

\bibliography{bibliography}

\begin{thebibliography}{91}%
\makeatletter
\providecommand \@ifxundefined [1]{%
 \@ifx{#1\undefined}
}%
\providecommand \@ifnum [1]{%
 \ifnum #1\expandafter \@firstoftwo
 \else \expandafter \@secondoftwo
 \fi
}%
\providecommand \@ifx [1]{%
 \ifx #1\expandafter \@firstoftwo
 \else \expandafter \@secondoftwo
 \fi
}%
\providecommand \natexlab [1]{#1}%
\providecommand \enquote  [1]{``#1''}%
\providecommand \bibnamefont  [1]{#1}%
\providecommand \bibfnamefont [1]{#1}%
\providecommand \citenamefont [1]{#1}%
\providecommand \href@noop [0]{\@secondoftwo}%
\providecommand \href [0]{\begingroup \@sanitize@url \@href}%
\providecommand \@href[1]{\@@startlink{#1}\@@href}%
\providecommand \@@href[1]{\endgroup#1\@@endlink}%
\providecommand \@sanitize@url [0]{\catcode `\\12\catcode `\$12\catcode
  `\&12\catcode `\#12\catcode `\^12\catcode `\_12\catcode `\%12\relax}%
\providecommand \@@startlink[1]{}%
\providecommand \@@endlink[0]{}%
\providecommand \url  [0]{\begingroup\@sanitize@url \@url }%
\providecommand \@url [1]{\endgroup\@href {#1}{\urlprefix }}%
\providecommand \urlprefix  [0]{URL }%
\providecommand \Eprint [0]{\href }%
\providecommand \doibase [0]{http://dx.doi.org/}%
\providecommand \selectlanguage [0]{\@gobble}%
\providecommand \bibinfo  [0]{\@secondoftwo}%
\providecommand \bibfield  [0]{\@secondoftwo}%
\providecommand \translation [1]{[#1]}%
\providecommand \BibitemOpen [0]{}%
\providecommand \bibitemStop [0]{}%
\providecommand \bibitemNoStop [0]{.\EOS\space}%
\providecommand \EOS [0]{\spacefactor3000\relax}%
\providecommand \BibitemShut  [1]{\csname bibitem#1\endcsname}%
\let\auto@bib@innerbib\@empty
\bibitem [{\citenamefont {{Alfv{\'e}n}}(1942)}]{alfven42}%
  \BibitemOpen
  \bibfield  {author} {\bibinfo {author} {\bibfnamefont {H.}~\bibnamefont
  {{Alfv{\'e}n}}},\ }\href {\doibase 10.1038/150405d0} {\bibfield  {journal}
  {\bibinfo  {journal} {\nat}\ }\textbf {\bibinfo {volume} {150}},\ \bibinfo
  {pages} {405} (\bibinfo {year} {1942})}\BibitemShut {NoStop}%
\bibitem [{\citenamefont {{Cranmer}}\ and\ \citenamefont
  {{Winebarger}}(2019)}]{cranmer19}%
  \BibitemOpen
  \bibfield  {author} {\bibinfo {author} {\bibfnamefont {S.~R.}\ \bibnamefont
  {{Cranmer}}}\ and\ \bibinfo {author} {\bibfnamefont {A.~R.}\ \bibnamefont
  {{Winebarger}}},\ }\href {\doibase 10.1146/annurev-astro-091918-104416}
  {\bibfield  {journal} {\bibinfo  {journal} {\araa}\ }\textbf {\bibinfo
  {volume} {57}},\ \bibinfo {pages} {157} (\bibinfo {year} {2019})}\BibitemShut
  {NoStop}%
\bibitem [{\citenamefont {{Chen}}(2022)}]{chen22}%
  \BibitemOpen
  \bibfield  {author} {\bibinfo {author} {\bibfnamefont {C.~H.~K.}\
  \bibnamefont {{Chen}}},\ }\href {\doibase 10.1038/s41550-022-01716-w}
  {\bibfield  {journal} {\bibinfo  {journal} {\natastro}\ }\textbf {\bibinfo
  {volume} {6}},\ \bibinfo {pages} {637} (\bibinfo {year} {2022})}\BibitemShut
  {NoStop}%
\bibitem [{\citenamefont {{Balbus}}\ and\ \citenamefont
  {{Hawley}}(1998)}]{balbus98}%
  \BibitemOpen
  \bibfield  {author} {\bibinfo {author} {\bibfnamefont {S.~A.}\ \bibnamefont
  {{Balbus}}}\ and\ \bibinfo {author} {\bibfnamefont {J.~F.}\ \bibnamefont
  {{Hawley}}},\ }\href {\doibase 10.1103/RevModPhys.70.1} {\bibfield  {journal}
  {\bibinfo  {journal} {\rmp}\ }\textbf {\bibinfo {volume} {70}},\ \bibinfo
  {pages} {1} (\bibinfo {year} {1998})}\BibitemShut {NoStop}%
\bibitem [{\citenamefont {{McKee}}\ and\ \citenamefont
  {{Ostriker}}(2007)}]{mckee07}%
  \BibitemOpen
  \bibfield  {author} {\bibinfo {author} {\bibfnamefont {C.~F.}\ \bibnamefont
  {{McKee}}}\ and\ \bibinfo {author} {\bibfnamefont {E.~C.}\ \bibnamefont
  {{Ostriker}}},\ }\href {\doibase 10.1146/annurev.astro.45.051806.110602}
  {\bibfield  {journal} {\bibinfo  {journal} {\araa}\ }\textbf {\bibinfo
  {volume} {45}},\ \bibinfo {pages} {565} (\bibinfo {year} {2007})}\BibitemShut
  {NoStop}%
\bibitem [{\citenamefont {{Kulsrud}}\ and\ \citenamefont
  {{Zweibel}}(2008)}]{kulsrud08}%
  \BibitemOpen
  \bibfield  {author} {\bibinfo {author} {\bibfnamefont {R.~M.}\ \bibnamefont
  {{Kulsrud}}}\ and\ \bibinfo {author} {\bibfnamefont {E.~G.}\ \bibnamefont
  {{Zweibel}}},\ }\href {\doibase 10.1088/0034-4885/71/4/046901} {\bibfield
  {journal} {\bibinfo  {journal} {\reppp}\ }\textbf {\bibinfo {volume} {71}},\
  \bibinfo {eid} {046901} (\bibinfo {year} {2008})}\BibitemShut {NoStop}%
\bibitem [{\citenamefont {{Zhuravleva}}\ \emph {et~al.}(2014)\citenamefont
  {{Zhuravleva}}, \citenamefont {{Churazov}}, \citenamefont {{Schekochihin}},
  \citenamefont {{Allen}}, \citenamefont {{Ar{\'e}valo}}, \citenamefont
  {{Fabian}}, \citenamefont {{Forman}}, \citenamefont {{Sanders}},
  \citenamefont {{Simionescu}}, \citenamefont {{Sunyaev}}, \citenamefont
  {{Vikhlinin}},\ and\ \citenamefont {{Werner}}}]{zhuravleva14}%
  \BibitemOpen
  \bibfield  {author} {\bibinfo {author} {\bibfnamefont {I.}~\bibnamefont
  {{Zhuravleva}}}, \bibinfo {author} {\bibfnamefont {E.}~\bibnamefont
  {{Churazov}}}, \bibinfo {author} {\bibfnamefont {A.~A.}\ \bibnamefont
  {{Schekochihin}}}, \bibinfo {author} {\bibfnamefont {S.~W.}\ \bibnamefont
  {{Allen}}}, \bibinfo {author} {\bibfnamefont {P.}~\bibnamefont
  {{Ar{\'e}valo}}}, \bibinfo {author} {\bibfnamefont {A.~C.}\ \bibnamefont
  {{Fabian}}}, \bibinfo {author} {\bibfnamefont {W.~R.}\ \bibnamefont
  {{Forman}}}, \bibinfo {author} {\bibfnamefont {J.~S.}\ \bibnamefont
  {{Sanders}}}, \bibinfo {author} {\bibfnamefont {A.}~\bibnamefont
  {{Simionescu}}}, \bibinfo {author} {\bibfnamefont {R.}~\bibnamefont
  {{Sunyaev}}}, \bibinfo {author} {\bibfnamefont {A.}~\bibnamefont
  {{Vikhlinin}}}, \ and\ \bibinfo {author} {\bibfnamefont {N.}~\bibnamefont
  {{Werner}}},\ }\href {\doibase 10.1038/nature13830} {\bibfield  {journal}
  {\bibinfo  {journal} {\nat}\ }\textbf {\bibinfo {volume} {515}},\ \bibinfo
  {pages} {85} (\bibinfo {year} {2014})}\BibitemShut {NoStop}%
\bibitem [{\citenamefont {{Bruno}}\ and\ \citenamefont
  {{Carbone}}(2013)}]{bruno13}%
  \BibitemOpen
  \bibfield  {author} {\bibinfo {author} {\bibfnamefont {R.}~\bibnamefont
  {{Bruno}}}\ and\ \bibinfo {author} {\bibfnamefont {V.}~\bibnamefont
  {{Carbone}}},\ }\href {\doibase 10.12942/lrsp-2013-2} {\bibfield  {journal}
  {\bibinfo  {journal} {\lrsp}\ }\textbf {\bibinfo {volume} {10}},\ \bibinfo
  {pages} {2} (\bibinfo {year} {2013})}\BibitemShut {NoStop}%
\bibitem [{\citenamefont {{Alexandrova}}\ \emph {et~al.}(2013)\citenamefont
  {{Alexandrova}}, \citenamefont {{Chen}}, \citenamefont {{Sorriso-Valvo}},
  \citenamefont {{Horbury}},\ and\ \citenamefont {{Bale}}}]{alexandrova13a}%
  \BibitemOpen
  \bibfield  {author} {\bibinfo {author} {\bibfnamefont {O.}~\bibnamefont
  {{Alexandrova}}}, \bibinfo {author} {\bibfnamefont {C.~H.~K.}\ \bibnamefont
  {{Chen}}}, \bibinfo {author} {\bibfnamefont {L.}~\bibnamefont
  {{Sorriso-Valvo}}}, \bibinfo {author} {\bibfnamefont {T.~S.}\ \bibnamefont
  {{Horbury}}}, \ and\ \bibinfo {author} {\bibfnamefont {S.~D.}\ \bibnamefont
  {{Bale}}},\ }\href {\doibase 10.1007/s11214-013-0004-8} {\bibfield  {journal}
  {\bibinfo  {journal} {\ssr}\ }\textbf {\bibinfo {volume} {178}},\ \bibinfo
  {pages} {101} (\bibinfo {year} {2013})}\BibitemShut {NoStop}%
\bibitem [{\citenamefont {{Chen}}(2016)}]{chen16b}%
  \BibitemOpen
  \bibfield  {author} {\bibinfo {author} {\bibfnamefont {C.~H.~K.}\
  \bibnamefont {{Chen}}},\ }\href@noop {} {\bibfield  {journal} {\bibinfo
  {journal} {\jpp}\ }\textbf {\bibinfo {volume} {82}},\ \bibinfo {pages}
  {535820602} (\bibinfo {year} {2016})}\BibitemShut {NoStop}%
\bibitem [{\citenamefont {{Wilson III}}\ \emph {et~al.}(2021)\citenamefont
  {{Wilson III}}, \citenamefont {{Brosius}}, \citenamefont {{Gopalswamy}},
  \citenamefont {{Nieves-Chinchilla}}, \citenamefont {{Szabo}}, \citenamefont
  {{Hurley}}, \citenamefont {{Phan}}, \citenamefont {{Kasper}}, \citenamefont
  {{Lugaz}}, \citenamefont {{Richardson}}, \citenamefont {{Chen}},
  \citenamefont {{Verscharen}}, \citenamefont {{Wicks}},\ and\ \citenamefont
  {{TenBarge}}}]{wilson21}%
  \BibitemOpen
  \bibfield  {author} {\bibinfo {author} {\bibfnamefont {L.~B.}\ \bibnamefont
  {{Wilson III}}}, \bibinfo {author} {\bibfnamefont {A.~L.}\ \bibnamefont
  {{Brosius}}}, \bibinfo {author} {\bibfnamefont {N.}~\bibnamefont
  {{Gopalswamy}}}, \bibinfo {author} {\bibfnamefont {T.}~\bibnamefont
  {{Nieves-Chinchilla}}}, \bibinfo {author} {\bibfnamefont {A.}~\bibnamefont
  {{Szabo}}}, \bibinfo {author} {\bibfnamefont {K.}~\bibnamefont {{Hurley}}},
  \bibinfo {author} {\bibfnamefont {T.}~\bibnamefont {{Phan}}}, \bibinfo
  {author} {\bibfnamefont {J.~C.}\ \bibnamefont {{Kasper}}}, \bibinfo {author}
  {\bibfnamefont {N.}~\bibnamefont {{Lugaz}}}, \bibinfo {author} {\bibfnamefont
  {I.~G.}\ \bibnamefont {{Richardson}}}, \bibinfo {author} {\bibfnamefont
  {C.~H.~K.}\ \bibnamefont {{Chen}}}, \bibinfo {author} {\bibfnamefont
  {D.}~\bibnamefont {{Verscharen}}}, \bibinfo {author} {\bibfnamefont {R.~T.}\
  \bibnamefont {{Wicks}}}, \ and\ \bibinfo {author} {\bibfnamefont {J.~M.}\
  \bibnamefont {{TenBarge}}},\ }\href {\doibase
  10.1029/2020RG00071410.1002/essoar.10504309.2} {\bibfield  {journal}
  {\bibinfo  {journal} {\rog}\ }\textbf {\bibinfo {volume} {59}},\ \bibinfo
  {pages} {e2020RG000714} (\bibinfo {year} {2021})}\BibitemShut {NoStop}%
\bibitem [{\citenamefont {{Schekochihin}}(2022)}]{schekochihin22}%
  \BibitemOpen
  \bibfield  {author} {\bibinfo {author} {\bibfnamefont {A.~A.}\ \bibnamefont
  {{Schekochihin}}},\ }\href {\doibase 10.1017/S0022377822000721} {\bibfield
  {journal} {\bibinfo  {journal} {\jpp}\ }\textbf {\bibinfo {volume} {88}},\
  \bibinfo {eid} {155880501} (\bibinfo {year} {2022})}\BibitemShut {NoStop}%
\bibitem [{\citenamefont {{Iroshnikov}}(1963)}]{iroshnikov63t}%
  \BibitemOpen
  \bibfield  {author} {\bibinfo {author} {\bibfnamefont {P.~S.}\ \bibnamefont
  {{Iroshnikov}}},\ }\href@noop {} {\bibfield  {journal} {\bibinfo  {journal}
  {Astron. Zh.}\ }\textbf {\bibinfo {volume} {40}},\ \bibinfo {pages} {742}
  (\bibinfo {year} {1963})},\ \bibinfo {note} {[Sov. Astron. \textbf{7}, 566
  (1964)]}\BibitemShut {NoStop}%
\bibitem [{\citenamefont {{Kraichnan}}(1965)}]{kraichnan65}%
  \BibitemOpen
  \bibfield  {author} {\bibinfo {author} {\bibfnamefont {R.~H.}\ \bibnamefont
  {{Kraichnan}}},\ }\href {\doibase 10.1063/1.1761412} {\bibfield  {journal}
  {\bibinfo  {journal} {\pof}\ }\textbf {\bibinfo {volume} {8}},\ \bibinfo
  {pages} {1385} (\bibinfo {year} {1965})}\BibitemShut {NoStop}%
\bibitem [{\citenamefont {{Elsasser}}(1950)}]{elsasser50}%
  \BibitemOpen
  \bibfield  {author} {\bibinfo {author} {\bibfnamefont {W.~M.}\ \bibnamefont
  {{Elsasser}}},\ }\href {\doibase 10.1103/PhysRev.79.183} {\bibfield
  {journal} {\bibinfo  {journal} {\pr}\ }\textbf {\bibinfo {volume} {79}},\
  \bibinfo {pages} {183} (\bibinfo {year} {1950})}\BibitemShut {NoStop}%
\bibitem [{\citenamefont {{Shebalin}}\ \emph {et~al.}(1983)\citenamefont
  {{Shebalin}}, \citenamefont {{Matthaeus}},\ and\ \citenamefont
  {{Montgomery}}}]{shebalin83}%
  \BibitemOpen
  \bibfield  {author} {\bibinfo {author} {\bibfnamefont {J.~V.}\ \bibnamefont
  {{Shebalin}}}, \bibinfo {author} {\bibfnamefont {W.~H.}\ \bibnamefont
  {{Matthaeus}}}, \ and\ \bibinfo {author} {\bibfnamefont {D.}~\bibnamefont
  {{Montgomery}}},\ }\href {\doibase 10.1017/S0022377800000933} {\bibfield
  {journal} {\bibinfo  {journal} {\jpp}\ }\textbf {\bibinfo {volume} {29}},\
  \bibinfo {pages} {525} (\bibinfo {year} {1983})}\BibitemShut {NoStop}%
\bibitem [{\citenamefont {{Meyrand}}\ \emph {et~al.}(2016)\citenamefont
  {{Meyrand}}, \citenamefont {{Galtier}},\ and\ \citenamefont
  {{Kiyani}}}]{meyrand16}%
  \BibitemOpen
  \bibfield  {author} {\bibinfo {author} {\bibfnamefont {R.}~\bibnamefont
  {{Meyrand}}}, \bibinfo {author} {\bibfnamefont {S.}~\bibnamefont
  {{Galtier}}}, \ and\ \bibinfo {author} {\bibfnamefont {K.~H.}\ \bibnamefont
  {{Kiyani}}},\ }\href {\doibase 10.1103/PhysRevLett.116.105002} {\bibfield
  {journal} {\bibinfo  {journal} {\prl}\ }\textbf {\bibinfo {volume} {116}},\
  \bibinfo {eid} {105002} (\bibinfo {year} {2016})}\BibitemShut {NoStop}%
\bibitem [{\citenamefont {{Sridhar}}\ and\ \citenamefont
  {{Goldreich}}(1994)}]{sridhar94}%
  \BibitemOpen
  \bibfield  {author} {\bibinfo {author} {\bibfnamefont {S.}~\bibnamefont
  {{Sridhar}}}\ and\ \bibinfo {author} {\bibfnamefont {P.}~\bibnamefont
  {{Goldreich}}},\ }\href {\doibase 10.1086/174600} {\bibfield  {journal}
  {\bibinfo  {journal} {\apj}\ }\textbf {\bibinfo {volume} {432}},\ \bibinfo
  {pages} {612} (\bibinfo {year} {1994})}\BibitemShut {NoStop}%
\bibitem [{\citenamefont {{Goldreich}}\ and\ \citenamefont
  {{Sridhar}}(1995)}]{goldreich95}%
  \BibitemOpen
  \bibfield  {author} {\bibinfo {author} {\bibfnamefont {P.}~\bibnamefont
  {{Goldreich}}}\ and\ \bibinfo {author} {\bibfnamefont {S.}~\bibnamefont
  {{Sridhar}}},\ }\href {\doibase 10.1086/175121} {\bibfield  {journal}
  {\bibinfo  {journal} {\apj}\ }\textbf {\bibinfo {volume} {438}},\ \bibinfo
  {pages} {763} (\bibinfo {year} {1995})}\BibitemShut {NoStop}%
\bibitem [{\citenamefont {{Montgomery}}\ and\ \citenamefont
  {{Matthaeus}}(1995)}]{montgomery95}%
  \BibitemOpen
  \bibfield  {author} {\bibinfo {author} {\bibfnamefont {D.}~\bibnamefont
  {{Montgomery}}}\ and\ \bibinfo {author} {\bibfnamefont {W.~H.}\ \bibnamefont
  {{Matthaeus}}},\ }\href {\doibase 10.1086/175910} {\bibfield  {journal}
  {\bibinfo  {journal} {\apj}\ }\textbf {\bibinfo {volume} {447}},\ \bibinfo
  {pages} {706} (\bibinfo {year} {1995})}\BibitemShut {NoStop}%
\bibitem [{\citenamefont {{Ng}}\ and\ \citenamefont
  {{Bhattacharjee}}(1996)}]{ng96}%
  \BibitemOpen
  \bibfield  {author} {\bibinfo {author} {\bibfnamefont {C.~S.}\ \bibnamefont
  {{Ng}}}\ and\ \bibinfo {author} {\bibfnamefont {A.}~\bibnamefont
  {{Bhattacharjee}}},\ }\href {\doibase 10.1086/177468} {\bibfield  {journal}
  {\bibinfo  {journal} {\apj}\ }\textbf {\bibinfo {volume} {465}},\ \bibinfo
  {pages} {845} (\bibinfo {year} {1996})}\BibitemShut {NoStop}%
\bibitem [{\citenamefont {{Ng}}\ and\ \citenamefont
  {{Bhattacharjee}}(1997)}]{ng97}%
  \BibitemOpen
  \bibfield  {author} {\bibinfo {author} {\bibfnamefont {C.~S.}\ \bibnamefont
  {{Ng}}}\ and\ \bibinfo {author} {\bibfnamefont {A.}~\bibnamefont
  {{Bhattacharjee}}},\ }\href {\doibase 10.1063/1.872158} {\bibfield  {journal}
  {\bibinfo  {journal} {\pop}\ }\textbf {\bibinfo {volume} {4}},\ \bibinfo
  {pages} {605} (\bibinfo {year} {1997})}\BibitemShut {NoStop}%
\bibitem [{\citenamefont {{Goldreich}}\ and\ \citenamefont
  {{Sridhar}}(1997)}]{goldreich97}%
  \BibitemOpen
  \bibfield  {author} {\bibinfo {author} {\bibfnamefont {P.}~\bibnamefont
  {{Goldreich}}}\ and\ \bibinfo {author} {\bibfnamefont {S.}~\bibnamefont
  {{Sridhar}}},\ }\href {\doibase 10.1086/304442} {\bibfield  {journal}
  {\bibinfo  {journal} {\apj}\ }\textbf {\bibinfo {volume} {485}},\ \bibinfo
  {pages} {680} (\bibinfo {year} {1997})}\BibitemShut {NoStop}%
\bibitem [{\citenamefont {{Galtier}}\ \emph {et~al.}(2000)\citenamefont
  {{Galtier}}, \citenamefont {{Nazarenko}}, \citenamefont {{Newell}},\ and\
  \citenamefont {{Pouquet}}}]{galtier00}%
  \BibitemOpen
  \bibfield  {author} {\bibinfo {author} {\bibfnamefont {S.}~\bibnamefont
  {{Galtier}}}, \bibinfo {author} {\bibfnamefont {S.~V.}\ \bibnamefont
  {{Nazarenko}}}, \bibinfo {author} {\bibfnamefont {A.~C.}\ \bibnamefont
  {{Newell}}}, \ and\ \bibinfo {author} {\bibfnamefont {A.}~\bibnamefont
  {{Pouquet}}},\ }\href {\doibase 10.1017/S0022377899008284} {\bibfield
  {journal} {\bibinfo  {journal} {\jpp}\ }\textbf {\bibinfo {volume} {63}},\
  \bibinfo {pages} {447} (\bibinfo {year} {2000})}\BibitemShut {NoStop}%
\bibitem [{\citenamefont {{Galtier}}\ \emph {et~al.}(2002)\citenamefont
  {{Galtier}}, \citenamefont {{Nazarenko}}, \citenamefont {{Newell}},\ and\
  \citenamefont {{Pouquet}}}]{galtier02}%
  \BibitemOpen
  \bibfield  {author} {\bibinfo {author} {\bibfnamefont {S.}~\bibnamefont
  {{Galtier}}}, \bibinfo {author} {\bibfnamefont {S.~V.}\ \bibnamefont
  {{Nazarenko}}}, \bibinfo {author} {\bibfnamefont {A.~C.}\ \bibnamefont
  {{Newell}}}, \ and\ \bibinfo {author} {\bibfnamefont {A.}~\bibnamefont
  {{Pouquet}}},\ }\href {\doibase 10.1086/338791} {\bibfield  {journal}
  {\bibinfo  {journal} {\apjl}\ }\textbf {\bibinfo {volume} {564}},\ \bibinfo
  {pages} {L49} (\bibinfo {year} {2002})}\BibitemShut {NoStop}%
\bibitem [{\citenamefont {{Lithwick}}\ and\ \citenamefont
  {{Goldreich}}(2003)}]{lithwick03}%
  \BibitemOpen
  \bibfield  {author} {\bibinfo {author} {\bibfnamefont {Y.}~\bibnamefont
  {{Lithwick}}}\ and\ \bibinfo {author} {\bibfnamefont {P.}~\bibnamefont
  {{Goldreich}}},\ }\href {\doibase 10.1086/344676} {\bibfield  {journal}
  {\bibinfo  {journal} {\apj}\ }\textbf {\bibinfo {volume} {582}},\ \bibinfo
  {pages} {1220} (\bibinfo {year} {2003})}\BibitemShut {NoStop}%
\bibitem [{\citenamefont {{Boldyrev}}\ and\ \citenamefont
  {{Perez}}(2009)}]{boldyrev09a}%
  \BibitemOpen
  \bibfield  {author} {\bibinfo {author} {\bibfnamefont {S.}~\bibnamefont
  {{Boldyrev}}}\ and\ \bibinfo {author} {\bibfnamefont {J.~C.}\ \bibnamefont
  {{Perez}}},\ }\href {\doibase 10.1103/PhysRevLett.103.225001} {\bibfield
  {journal} {\bibinfo  {journal} {\prl}\ }\textbf {\bibinfo {volume} {103}},\
  \bibinfo {eid} {225001} (\bibinfo {year} {2009})}\BibitemShut {NoStop}%
\bibitem [{\citenamefont {{Schekochihin}}\ \emph {et~al.}(2012)\citenamefont
  {{Schekochihin}}, \citenamefont {{Nazarenko}},\ and\ \citenamefont
  {{Yousef}}}]{schekochihin12}%
  \BibitemOpen
  \bibfield  {author} {\bibinfo {author} {\bibfnamefont {A.~A.}\ \bibnamefont
  {{Schekochihin}}}, \bibinfo {author} {\bibfnamefont {S.~V.}\ \bibnamefont
  {{Nazarenko}}}, \ and\ \bibinfo {author} {\bibfnamefont {T.~A.}\ \bibnamefont
  {{Yousef}}},\ }\href {\doibase 10.1103/PhysRevE.85.036406} {\bibfield
  {journal} {\bibinfo  {journal} {\pre}\ }\textbf {\bibinfo {volume} {85}},\
  \bibinfo {eid} {036406} (\bibinfo {year} {2012})}\BibitemShut {NoStop}%
\bibitem [{\citenamefont {{Perez}}\ and\ \citenamefont
  {{Boldyrev}}(2008)}]{perez08}%
  \BibitemOpen
  \bibfield  {author} {\bibinfo {author} {\bibfnamefont {J.~C.}\ \bibnamefont
  {{Perez}}}\ and\ \bibinfo {author} {\bibfnamefont {S.}~\bibnamefont
  {{Boldyrev}}},\ }\href {\doibase 10.1086/526342} {\bibfield  {journal}
  {\bibinfo  {journal} {\apjl}\ }\textbf {\bibinfo {volume} {672}},\ \bibinfo
  {pages} {L61} (\bibinfo {year} {2008})}\BibitemShut {NoStop}%
\bibitem [{\citenamefont {{Meyrand}}\ \emph {et~al.}(2015)\citenamefont
  {{Meyrand}}, \citenamefont {{Kiyani}},\ and\ \citenamefont
  {{Galtier}}}]{meyrand15}%
  \BibitemOpen
  \bibfield  {author} {\bibinfo {author} {\bibfnamefont {R.}~\bibnamefont
  {{Meyrand}}}, \bibinfo {author} {\bibfnamefont {K.~H.}\ \bibnamefont
  {{Kiyani}}}, \ and\ \bibinfo {author} {\bibfnamefont {S.}~\bibnamefont
  {{Galtier}}},\ }\href {\doibase 10.1017/jfm.2015.141} {\bibfield  {journal}
  {\bibinfo  {journal} {\jfm}\ }\textbf {\bibinfo {volume} {770}},\ \bibinfo
  {eid} {R1} (\bibinfo {year} {2015})}\BibitemShut {NoStop}%
\bibitem [{\citenamefont {{Howes}}\ and\ \citenamefont
  {{Nielson}}(2013)}]{howes13a}%
  \BibitemOpen
  \bibfield  {author} {\bibinfo {author} {\bibfnamefont {G.~G.}\ \bibnamefont
  {{Howes}}}\ and\ \bibinfo {author} {\bibfnamefont {K.~D.}\ \bibnamefont
  {{Nielson}}},\ }\href {\doibase 10.1063/1.4812805} {\bibfield  {journal}
  {\bibinfo  {journal} {Physics of Plasmas}\ }\textbf {\bibinfo {volume}
  {20}},\ \bibinfo {eid} {072302} (\bibinfo {year} {2013})}\BibitemShut
  {NoStop}%
\bibitem [{\citenamefont {{Dorfman}}\ \emph {et~al.}(2025)\citenamefont
  {{Dorfman}}, \citenamefont {{Abler}}, \citenamefont {{Boldyrev}},
  \citenamefont {{Chen}},\ and\ \citenamefont {{Greess}}}]{dorfman25}%
  \BibitemOpen
  \bibfield  {author} {\bibinfo {author} {\bibfnamefont {S.}~\bibnamefont
  {{Dorfman}}}, \bibinfo {author} {\bibfnamefont {M.}~\bibnamefont {{Abler}}},
  \bibinfo {author} {\bibfnamefont {S.}~\bibnamefont {{Boldyrev}}}, \bibinfo
  {author} {\bibfnamefont {C.~H.~K.}\ \bibnamefont {{Chen}}}, \ and\ \bibinfo
  {author} {\bibfnamefont {S.}~\bibnamefont {{Greess}}},\ }\href {\doibase
  10.3847/1538-4357/ad9012} {\bibfield  {journal} {\bibinfo  {journal} {\apj}\
  }\textbf {\bibinfo {volume} {979}},\ \bibinfo {eid} {163} (\bibinfo {year}
  {2025})}\BibitemShut {NoStop}%
\bibitem [{\citenamefont {{Boldyrev}}(2006)}]{boldyrev06}%
  \BibitemOpen
  \bibfield  {author} {\bibinfo {author} {\bibfnamefont {S.}~\bibnamefont
  {{Boldyrev}}},\ }\href {\doibase 10.1103/PhysRevLett.96.115002} {\bibfield
  {journal} {\bibinfo  {journal} {\prl}\ }\textbf {\bibinfo {volume} {96}},\
  \bibinfo {pages} {115002} (\bibinfo {year} {2006})}\BibitemShut {NoStop}%
\bibitem [{\citenamefont {{Chandran}}\ \emph {et~al.}(2015)\citenamefont
  {{Chandran}}, \citenamefont {{Schekochihin}},\ and\ \citenamefont
  {{Mallet}}}]{chandran15}%
  \BibitemOpen
  \bibfield  {author} {\bibinfo {author} {\bibfnamefont {B.~D.~G.}\
  \bibnamefont {{Chandran}}}, \bibinfo {author} {\bibfnamefont {A.~A.}\
  \bibnamefont {{Schekochihin}}}, \ and\ \bibinfo {author} {\bibfnamefont
  {A.}~\bibnamefont {{Mallet}}},\ }\href {\doibase 10.1088/0004-637X/807/1/39}
  {\bibfield  {journal} {\bibinfo  {journal} {\apj}\ }\textbf {\bibinfo
  {volume} {807}},\ \bibinfo {eid} {39} (\bibinfo {year} {2015})}\BibitemShut
  {NoStop}%
\bibitem [{\citenamefont {{Mallet}}\ and\ \citenamefont
  {{Schekochihin}}(2017)}]{mallet17}%
  \BibitemOpen
  \bibfield  {author} {\bibinfo {author} {\bibfnamefont {A.}~\bibnamefont
  {{Mallet}}}\ and\ \bibinfo {author} {\bibfnamefont {A.~A.}\ \bibnamefont
  {{Schekochihin}}},\ }\href {\doibase 10.1093/mnras/stw3251} {\bibfield
  {journal} {\bibinfo  {journal} {\mnras}\ }\textbf {\bibinfo {volume} {466}},\
  \bibinfo {pages} {3918} (\bibinfo {year} {2017})}\BibitemShut {NoStop}%
\bibitem [{\citenamefont {{Armstrong}}\ \emph {et~al.}(1995)\citenamefont
  {{Armstrong}}, \citenamefont {{Rickett}},\ and\ \citenamefont
  {{Spangler}}}]{armstrong95}%
  \BibitemOpen
  \bibfield  {author} {\bibinfo {author} {\bibfnamefont {J.~W.}\ \bibnamefont
  {{Armstrong}}}, \bibinfo {author} {\bibfnamefont {B.~J.}\ \bibnamefont
  {{Rickett}}}, \ and\ \bibinfo {author} {\bibfnamefont {S.~R.}\ \bibnamefont
  {{Spangler}}},\ }\href {\doibase 10.1086/175515} {\bibfield  {journal}
  {\bibinfo  {journal} {\apj}\ }\textbf {\bibinfo {volume} {443}},\ \bibinfo
  {pages} {209} (\bibinfo {year} {1995})}\BibitemShut {NoStop}%
\bibitem [{\citenamefont {{Lithwick}}\ \emph {et~al.}(2007)\citenamefont
  {{Lithwick}}, \citenamefont {{Goldreich}},\ and\ \citenamefont
  {{Sridhar}}}]{lithwick07}%
  \BibitemOpen
  \bibfield  {author} {\bibinfo {author} {\bibfnamefont {Y.}~\bibnamefont
  {{Lithwick}}}, \bibinfo {author} {\bibfnamefont {P.}~\bibnamefont
  {{Goldreich}}}, \ and\ \bibinfo {author} {\bibfnamefont {S.}~\bibnamefont
  {{Sridhar}}},\ }\href {\doibase 10.1086/509884} {\bibfield  {journal}
  {\bibinfo  {journal} {\apj}\ }\textbf {\bibinfo {volume} {655}},\ \bibinfo
  {pages} {269} (\bibinfo {year} {2007})}\BibitemShut {NoStop}%
\bibitem [{\citenamefont {{Beresnyak}}\ and\ \citenamefont
  {{Lazarian}}(2008)}]{beresnyak08}%
  \BibitemOpen
  \bibfield  {author} {\bibinfo {author} {\bibfnamefont {A.}~\bibnamefont
  {{Beresnyak}}}\ and\ \bibinfo {author} {\bibfnamefont {A.}~\bibnamefont
  {{Lazarian}}},\ }\href {\doibase 10.1086/589428} {\bibfield  {journal}
  {\bibinfo  {journal} {\apj}\ }\textbf {\bibinfo {volume} {682}},\ \bibinfo
  {pages} {1070} (\bibinfo {year} {2008})}\BibitemShut {NoStop}%
\bibitem [{\citenamefont {{Chandran}}(2008)}]{chandran08}%
  \BibitemOpen
  \bibfield  {author} {\bibinfo {author} {\bibfnamefont {B.~D.~G.}\
  \bibnamefont {{Chandran}}},\ }\href {\doibase 10.1086/589432} {\bibfield
  {journal} {\bibinfo  {journal} {\apj}\ }\textbf {\bibinfo {volume} {685}},\
  \bibinfo {pages} {646} (\bibinfo {year} {2008})}\BibitemShut {NoStop}%
\bibitem [{\citenamefont {{Perez}}\ and\ \citenamefont
  {{Boldyrev}}(2009)}]{perez09}%
  \BibitemOpen
  \bibfield  {author} {\bibinfo {author} {\bibfnamefont {J.~C.}\ \bibnamefont
  {{Perez}}}\ and\ \bibinfo {author} {\bibfnamefont {S.}~\bibnamefont
  {{Boldyrev}}},\ }\href {\doibase 10.1103/PhysRevLett.102.025003} {\bibfield
  {journal} {\bibinfo  {journal} {\prl}\ }\textbf {\bibinfo {volume} {102}},\
  \bibinfo {pages} {025003} (\bibinfo {year} {2009})}\BibitemShut {NoStop}%
\bibitem [{\citenamefont {{Podesta}}\ and\ \citenamefont
  {{Bhattacharjee}}(2010)}]{podesta10c}%
  \BibitemOpen
  \bibfield  {author} {\bibinfo {author} {\bibfnamefont {J.~J.}\ \bibnamefont
  {{Podesta}}}\ and\ \bibinfo {author} {\bibfnamefont {A.}~\bibnamefont
  {{Bhattacharjee}}},\ }\href {\doibase 10.1088/0004-637X/718/2/1151}
  {\bibfield  {journal} {\bibinfo  {journal} {\apj}\ }\textbf {\bibinfo
  {volume} {718}},\ \bibinfo {pages} {1151} (\bibinfo {year}
  {2010})}\BibitemShut {NoStop}%
\bibitem [{\citenamefont {{Chandran}}\ and\ \citenamefont
  {{Perez}}(2019)}]{chandran19}%
  \BibitemOpen
  \bibfield  {author} {\bibinfo {author} {\bibfnamefont {B.~D.~G.}\
  \bibnamefont {{Chandran}}}\ and\ \bibinfo {author} {\bibfnamefont {J.~C.}\
  \bibnamefont {{Perez}}},\ }\href {\doibase 10.1017/S0022377819000540}
  {\bibfield  {journal} {\bibinfo  {journal} {\jpp}\ }\textbf {\bibinfo
  {volume} {85}},\ \bibinfo {eid} {905850409} (\bibinfo {year}
  {2019})}\BibitemShut {NoStop}%
\bibitem [{\citenamefont {{Chandran}}\ \emph {et~al.}(2025)\citenamefont
  {{Chandran}}, \citenamefont {{Sioulas}}, \citenamefont {{Bale}},
  \citenamefont {{Bowen}}, \citenamefont {{David}}, \citenamefont {{Meyrand}},\
  and\ \citenamefont {{Yerger}}}]{chandran25}%
  \BibitemOpen
  \bibfield  {author} {\bibinfo {author} {\bibfnamefont {B.~D.~G.}\
  \bibnamefont {{Chandran}}}, \bibinfo {author} {\bibfnamefont
  {N.}~\bibnamefont {{Sioulas}}}, \bibinfo {author} {\bibfnamefont
  {S.}~\bibnamefont {{Bale}}}, \bibinfo {author} {\bibfnamefont
  {T.}~\bibnamefont {{Bowen}}}, \bibinfo {author} {\bibfnamefont
  {V.}~\bibnamefont {{David}}}, \bibinfo {author} {\bibfnamefont
  {R.}~\bibnamefont {{Meyrand}}}, \ and\ \bibinfo {author} {\bibfnamefont
  {E.}~\bibnamefont {{Yerger}}},\ }\href {\doibase 10.1017/S0022377825000194}
  {\bibfield  {journal} {\bibinfo  {journal} {\jpp}\ }\textbf {\bibinfo
  {volume} {91}},\ \bibinfo {eid} {E57} (\bibinfo {year} {2025})}\BibitemShut
  {NoStop}%
\bibitem [{\citenamefont {{Howes}}(2018)}]{howes18b}%
  \BibitemOpen
  \bibfield  {author} {\bibinfo {author} {\bibfnamefont {G.~G.}\ \bibnamefont
  {{Howes}}},\ }\href {\doibase 10.1063/1.5025421} {\bibfield  {journal}
  {\bibinfo  {journal} {\pop}\ }\textbf {\bibinfo {volume} {25}},\ \bibinfo
  {eid} {055501} (\bibinfo {year} {2018})}\BibitemShut {NoStop}%
\bibitem [{\citenamefont {{Robinson}}\ \emph {et~al.}(1968)\citenamefont
  {{Robinson}}, \citenamefont {{Rusbridge}},\ and\ \citenamefont
  {{Saunders}}}]{robinson68}%
  \BibitemOpen
  \bibfield  {author} {\bibinfo {author} {\bibfnamefont {D.~C.}\ \bibnamefont
  {{Robinson}}}, \bibinfo {author} {\bibfnamefont {M.~G.}\ \bibnamefont
  {{Rusbridge}}}, \ and\ \bibinfo {author} {\bibfnamefont {P.~A.~H.}\
  \bibnamefont {{Saunders}}},\ }\href {\doibase 10.1088/0032-1028/10/11/305}
  {\bibfield  {journal} {\bibinfo  {journal} {\pp}\ }\textbf {\bibinfo {volume}
  {10}},\ \bibinfo {pages} {1005} (\bibinfo {year} {1968})}\BibitemShut
  {NoStop}%
\bibitem [{\citenamefont {{Robinson}}\ and\ \citenamefont
  {{Rusbridge}}(1971)}]{robinson71}%
  \BibitemOpen
  \bibfield  {author} {\bibinfo {author} {\bibfnamefont {D.~C.}\ \bibnamefont
  {{Robinson}}}\ and\ \bibinfo {author} {\bibfnamefont {M.~G.}\ \bibnamefont
  {{Rusbridge}}},\ }\href {\doibase 10.1063/1.1693359} {\bibfield  {journal}
  {\bibinfo  {journal} {\pof}\ }\textbf {\bibinfo {volume} {14}},\ \bibinfo
  {pages} {2499} (\bibinfo {year} {1971})}\BibitemShut {NoStop}%
\bibitem [{\citenamefont {{Zweben}}\ \emph {et~al.}(1979)\citenamefont
  {{Zweben}}, \citenamefont {{Menyuk}},\ and\ \citenamefont
  {{Taylor}}}]{zweben79}%
  \BibitemOpen
  \bibfield  {author} {\bibinfo {author} {\bibfnamefont {S.~J.}\ \bibnamefont
  {{Zweben}}}, \bibinfo {author} {\bibfnamefont {C.~R.}\ \bibnamefont
  {{Menyuk}}}, \ and\ \bibinfo {author} {\bibfnamefont {R.~J.}\ \bibnamefont
  {{Taylor}}},\ }\href {\doibase 10.1103/PhysRevLett.42.1270} {\bibfield
  {journal} {\bibinfo  {journal} {\prl}\ }\textbf {\bibinfo {volume} {42}},\
  \bibinfo {pages} {1270} (\bibinfo {year} {1979})}\BibitemShut {NoStop}%
\bibitem [{\citenamefont {{Gekelman}}\ \emph {et~al.}(1994)\citenamefont
  {{Gekelman}}, \citenamefont {{Leneman}}, \citenamefont {{Maggs}},\ and\
  \citenamefont {{Vincena}}}]{gekelman94}%
  \BibitemOpen
  \bibfield  {author} {\bibinfo {author} {\bibfnamefont {W.}~\bibnamefont
  {{Gekelman}}}, \bibinfo {author} {\bibfnamefont {D.}~\bibnamefont
  {{Leneman}}}, \bibinfo {author} {\bibfnamefont {J.}~\bibnamefont {{Maggs}}},
  \ and\ \bibinfo {author} {\bibfnamefont {S.}~\bibnamefont {{Vincena}}},\
  }\href {\doibase 10.1063/1.870851} {\bibfield  {journal} {\bibinfo  {journal}
  {\pop}\ }\textbf {\bibinfo {volume} {1}},\ \bibinfo {pages} {3775} (\bibinfo
  {year} {1994})}\BibitemShut {NoStop}%
\bibitem [{\citenamefont {{Gekelman}}\ \emph {et~al.}(1997)\citenamefont
  {{Gekelman}}, \citenamefont {{Vincena}}, \citenamefont {{Leneman}},\ and\
  \citenamefont {{Maggs}}}]{gekelman97}%
  \BibitemOpen
  \bibfield  {author} {\bibinfo {author} {\bibfnamefont {W.}~\bibnamefont
  {{Gekelman}}}, \bibinfo {author} {\bibfnamefont {S.}~\bibnamefont
  {{Vincena}}}, \bibinfo {author} {\bibfnamefont {D.}~\bibnamefont
  {{Leneman}}}, \ and\ \bibinfo {author} {\bibfnamefont {J.}~\bibnamefont
  {{Maggs}}},\ }\href {\doibase 10.1029/96JA03683} {\bibfield  {journal}
  {\bibinfo  {journal} {\jgr}\ }\textbf {\bibinfo {volume} {102}},\ \bibinfo
  {pages} {7225} (\bibinfo {year} {1997})}\BibitemShut {NoStop}%
\bibitem [{\citenamefont {{Leneman}}\ \emph {et~al.}(1999)\citenamefont
  {{Leneman}}, \citenamefont {{Gekelman}},\ and\ \citenamefont
  {{Maggs}}}]{leneman99}%
  \BibitemOpen
  \bibfield  {author} {\bibinfo {author} {\bibfnamefont {D.}~\bibnamefont
  {{Leneman}}}, \bibinfo {author} {\bibfnamefont {W.}~\bibnamefont
  {{Gekelman}}}, \ and\ \bibinfo {author} {\bibfnamefont {J.}~\bibnamefont
  {{Maggs}}},\ }\href {\doibase 10.1103/PhysRevLett.82.2673} {\bibfield
  {journal} {\bibinfo  {journal} {\prl}\ }\textbf {\bibinfo {volume} {82}},\
  \bibinfo {pages} {2673} (\bibinfo {year} {1999})}\BibitemShut {NoStop}%
\bibitem [{\citenamefont {{Vincena}}\ \emph {et~al.}(2004)\citenamefont
  {{Vincena}}, \citenamefont {{Gekelman}},\ and\ \citenamefont
  {{Maggs}}}]{vincena04}%
  \BibitemOpen
  \bibfield  {author} {\bibinfo {author} {\bibfnamefont {S.}~\bibnamefont
  {{Vincena}}}, \bibinfo {author} {\bibfnamefont {W.}~\bibnamefont
  {{Gekelman}}}, \ and\ \bibinfo {author} {\bibfnamefont {J.}~\bibnamefont
  {{Maggs}}},\ }\href {\doibase 10.1103/PhysRevLett.93.105003} {\bibfield
  {journal} {\bibinfo  {journal} {\prl}\ }\textbf {\bibinfo {volume} {93}},\
  \bibinfo {eid} {105003} (\bibinfo {year} {2004})}\BibitemShut {NoStop}%
\bibitem [{\citenamefont {{Palmer}}\ \emph {et~al.}(2005)\citenamefont
  {{Palmer}}, \citenamefont {{Gekelman}},\ and\ \citenamefont
  {{Vincena}}}]{palmer05}%
  \BibitemOpen
  \bibfield  {author} {\bibinfo {author} {\bibfnamefont {N.}~\bibnamefont
  {{Palmer}}}, \bibinfo {author} {\bibfnamefont {W.}~\bibnamefont
  {{Gekelman}}}, \ and\ \bibinfo {author} {\bibfnamefont {S.}~\bibnamefont
  {{Vincena}}},\ }\href {\doibase 10.1063/1.1930796} {\bibfield  {journal}
  {\bibinfo  {journal} {\pop}\ }\textbf {\bibinfo {volume} {12}},\ \bibinfo
  {eid} {072102} (\bibinfo {year} {2005})}\BibitemShut {NoStop}%
\bibitem [{\citenamefont {{Kletzing}}\ \emph {et~al.}(2003)\citenamefont
  {{Kletzing}}, \citenamefont {{Bounds}}, \citenamefont {{Martin-Hiner}},
  \citenamefont {{Gekelman}},\ and\ \citenamefont {{Mitchell}}}]{kletzing03}%
  \BibitemOpen
  \bibfield  {author} {\bibinfo {author} {\bibfnamefont {C.~A.}\ \bibnamefont
  {{Kletzing}}}, \bibinfo {author} {\bibfnamefont {S.~R.}\ \bibnamefont
  {{Bounds}}}, \bibinfo {author} {\bibfnamefont {J.}~\bibnamefont
  {{Martin-Hiner}}}, \bibinfo {author} {\bibfnamefont {W.}~\bibnamefont
  {{Gekelman}}}, \ and\ \bibinfo {author} {\bibfnamefont {C.}~\bibnamefont
  {{Mitchell}}},\ }\href {\doibase 10.1103/PhysRevLett.90.035004} {\bibfield
  {journal} {\bibinfo  {journal} {\prl}\ }\textbf {\bibinfo {volume} {90}},\
  \bibinfo {eid} {035004} (\bibinfo {year} {2003})}\BibitemShut {NoStop}%
\bibitem [{\citenamefont {{Thuecks}}\ \emph {et~al.}(2009)\citenamefont
  {{Thuecks}}, \citenamefont {{Kletzing}}, \citenamefont {{Skiff}},
  \citenamefont {{Bounds}},\ and\ \citenamefont {{Vincena}}}]{thuecks09}%
  \BibitemOpen
  \bibfield  {author} {\bibinfo {author} {\bibfnamefont {D.~J.}\ \bibnamefont
  {{Thuecks}}}, \bibinfo {author} {\bibfnamefont {C.~A.}\ \bibnamefont
  {{Kletzing}}}, \bibinfo {author} {\bibfnamefont {F.}~\bibnamefont {{Skiff}}},
  \bibinfo {author} {\bibfnamefont {S.~R.}\ \bibnamefont {{Bounds}}}, \ and\
  \bibinfo {author} {\bibfnamefont {S.}~\bibnamefont {{Vincena}}},\ }\href
  {\doibase 10.1063/1.3140037} {\bibfield  {journal} {\bibinfo  {journal}
  {\pop}\ }\textbf {\bibinfo {volume} {16}},\ \bibinfo {eid} {052110} (\bibinfo
  {year} {2009})}\BibitemShut {NoStop}%
\bibitem [{\citenamefont {{Kletzing}}\ \emph {et~al.}(2010)\citenamefont
  {{Kletzing}}, \citenamefont {{Thuecks}}, \citenamefont {{Skiff}},
  \citenamefont {{Bounds}},\ and\ \citenamefont {{Vincena}}}]{kletzing10}%
  \BibitemOpen
  \bibfield  {author} {\bibinfo {author} {\bibfnamefont {C.~A.}\ \bibnamefont
  {{Kletzing}}}, \bibinfo {author} {\bibfnamefont {D.~J.}\ \bibnamefont
  {{Thuecks}}}, \bibinfo {author} {\bibfnamefont {F.}~\bibnamefont {{Skiff}}},
  \bibinfo {author} {\bibfnamefont {S.~R.}\ \bibnamefont {{Bounds}}}, \ and\
  \bibinfo {author} {\bibfnamefont {S.}~\bibnamefont {{Vincena}}},\ }\href
  {\doibase 10.1103/PhysRevLett.104.095001} {\bibfield  {journal} {\bibinfo
  {journal} {\prl}\ }\textbf {\bibinfo {volume} {104}},\ \bibinfo {eid}
  {095001} (\bibinfo {year} {2010})}\BibitemShut {NoStop}%
\bibitem [{\citenamefont {{Carter}}\ \emph {et~al.}(2006)\citenamefont
  {{Carter}}, \citenamefont {{Brugman}}, \citenamefont {{Pribyl}},\ and\
  \citenamefont {{Lybarger}}}]{carter06}%
  \BibitemOpen
  \bibfield  {author} {\bibinfo {author} {\bibfnamefont {T.~A.}\ \bibnamefont
  {{Carter}}}, \bibinfo {author} {\bibfnamefont {B.}~\bibnamefont {{Brugman}}},
  \bibinfo {author} {\bibfnamefont {P.}~\bibnamefont {{Pribyl}}}, \ and\
  \bibinfo {author} {\bibfnamefont {W.}~\bibnamefont {{Lybarger}}},\ }\href
  {\doibase 10.1103/PhysRevLett.96.155001} {\bibfield  {journal} {\bibinfo
  {journal} {\prl}\ }\textbf {\bibinfo {volume} {96}},\ \bibinfo {eid} {155001}
  (\bibinfo {year} {2006})}\BibitemShut {NoStop}%
\bibitem [{\citenamefont {{Brugman}}(2007)}]{brugman07}%
  \BibitemOpen
  \bibfield  {author} {\bibinfo {author} {\bibfnamefont {B.~T.}\ \bibnamefont
  {{Brugman}}},\ }\emph {\bibinfo {title} {{An investigation of the nonlinear
  interaction of Alfven waves}}},\ \href@noop {} {Ph.D. thesis},\ \bibinfo
  {school} {University of California, Los Angeles} (\bibinfo {year}
  {2007})\BibitemShut {NoStop}%
\bibitem [{\citenamefont {{Dorfman}}\ and\ \citenamefont
  {{Carter}}(2013)}]{dorfman13}%
  \BibitemOpen
  \bibfield  {author} {\bibinfo {author} {\bibfnamefont {S.}~\bibnamefont
  {{Dorfman}}}\ and\ \bibinfo {author} {\bibfnamefont {T.~A.}\ \bibnamefont
  {{Carter}}},\ }\href {\doibase 10.1103/PhysRevLett.110.195001} {\bibfield
  {journal} {\bibinfo  {journal} {\prl}\ }\textbf {\bibinfo {volume} {110}},\
  \bibinfo {eid} {195001} (\bibinfo {year} {2013})}\BibitemShut {NoStop}%
\bibitem [{\citenamefont {{Dorfman}}\ and\ \citenamefont
  {{Carter}}(2016)}]{dorfman16}%
  \BibitemOpen
  \bibfield  {author} {\bibinfo {author} {\bibfnamefont {S.}~\bibnamefont
  {{Dorfman}}}\ and\ \bibinfo {author} {\bibfnamefont {T.~A.}\ \bibnamefont
  {{Carter}}},\ }\href {\doibase 10.1103/PhysRevLett.116.195002} {\bibfield
  {journal} {\bibinfo  {journal} {\prl}\ }\textbf {\bibinfo {volume} {116}},\
  \bibinfo {eid} {195002} (\bibinfo {year} {2016})}\BibitemShut {NoStop}%
\bibitem [{\citenamefont {{Howes}}\ \emph {et~al.}(2012)\citenamefont
  {{Howes}}, \citenamefont {{Drake}}, \citenamefont {{Nielson}}, \citenamefont
  {{Carter}}, \citenamefont {{Kletzing}},\ and\ \citenamefont
  {{Skiff}}}]{howes12b}%
  \BibitemOpen
  \bibfield  {author} {\bibinfo {author} {\bibfnamefont {G.~G.}\ \bibnamefont
  {{Howes}}}, \bibinfo {author} {\bibfnamefont {D.~J.}\ \bibnamefont
  {{Drake}}}, \bibinfo {author} {\bibfnamefont {K.~D.}\ \bibnamefont
  {{Nielson}}}, \bibinfo {author} {\bibfnamefont {T.~A.}\ \bibnamefont
  {{Carter}}}, \bibinfo {author} {\bibfnamefont {C.~A.}\ \bibnamefont
  {{Kletzing}}}, \ and\ \bibinfo {author} {\bibfnamefont {F.}~\bibnamefont
  {{Skiff}}},\ }\href {\doibase 10.1103/PhysRevLett.109.255001} {\bibfield
  {journal} {\bibinfo  {journal} {\prl}\ }\textbf {\bibinfo {volume} {109}},\
  \bibinfo {eid} {255001} (\bibinfo {year} {2012})}\BibitemShut {NoStop}%
\bibitem [{\citenamefont {{Drake}}\ \emph {et~al.}(2013)\citenamefont
  {{Drake}}, \citenamefont {{Schroeder}}, \citenamefont {{Howes}},
  \citenamefont {{Kletzing}}, \citenamefont {{Skiff}}, \citenamefont
  {{Carter}},\ and\ \citenamefont {{Auerbach}}}]{drake13}%
  \BibitemOpen
  \bibfield  {author} {\bibinfo {author} {\bibfnamefont {D.~J.}\ \bibnamefont
  {{Drake}}}, \bibinfo {author} {\bibfnamefont {J.~W.~R.}\ \bibnamefont
  {{Schroeder}}}, \bibinfo {author} {\bibfnamefont {G.~G.}\ \bibnamefont
  {{Howes}}}, \bibinfo {author} {\bibfnamefont {C.~A.}\ \bibnamefont
  {{Kletzing}}}, \bibinfo {author} {\bibfnamefont {F.}~\bibnamefont {{Skiff}}},
  \bibinfo {author} {\bibfnamefont {T.~A.}\ \bibnamefont {{Carter}}}, \ and\
  \bibinfo {author} {\bibfnamefont {D.~W.}\ \bibnamefont {{Auerbach}}},\ }\href
  {\doibase 10.1063/1.4813242} {\bibfield  {journal} {\bibinfo  {journal}
  {\pop}\ }\textbf {\bibinfo {volume} {20}},\ \bibinfo {eid} {072901} (\bibinfo
  {year} {2013})}\BibitemShut {NoStop}%
\bibitem [{\citenamefont {{Howes}}\ \emph {et~al.}(2013)\citenamefont
  {{Howes}}, \citenamefont {{Nielson}}, \citenamefont {{Drake}}, \citenamefont
  {{Schroeder}}, \citenamefont {{Skiff}}, \citenamefont {{Kletzing}},\ and\
  \citenamefont {{Carter}}}]{howes13b}%
  \BibitemOpen
  \bibfield  {author} {\bibinfo {author} {\bibfnamefont {G.~G.}\ \bibnamefont
  {{Howes}}}, \bibinfo {author} {\bibfnamefont {K.~D.}\ \bibnamefont
  {{Nielson}}}, \bibinfo {author} {\bibfnamefont {D.~J.}\ \bibnamefont
  {{Drake}}}, \bibinfo {author} {\bibfnamefont {J.~W.~R.}\ \bibnamefont
  {{Schroeder}}}, \bibinfo {author} {\bibfnamefont {F.}~\bibnamefont
  {{Skiff}}}, \bibinfo {author} {\bibfnamefont {C.~A.}\ \bibnamefont
  {{Kletzing}}}, \ and\ \bibinfo {author} {\bibfnamefont {T.~A.}\ \bibnamefont
  {{Carter}}},\ }\href {\doibase 10.1063/1.4812808} {\bibfield  {journal}
  {\bibinfo  {journal} {\pop}\ }\textbf {\bibinfo {volume} {20}},\ \bibinfo
  {eid} {072304} (\bibinfo {year} {2013})}\BibitemShut {NoStop}%
\bibitem [{\citenamefont {{Drake}}\ \emph {et~al.}(2016)\citenamefont
  {{Drake}}, \citenamefont {{Howes}}, \citenamefont {{Rhudy}}, \citenamefont
  {{Terry}}, \citenamefont {{Carter}}, \citenamefont {{Kletzing}},
  \citenamefont {{Schroeder}},\ and\ \citenamefont {{Skiff}}}]{drake16}%
  \BibitemOpen
  \bibfield  {author} {\bibinfo {author} {\bibfnamefont {D.~J.}\ \bibnamefont
  {{Drake}}}, \bibinfo {author} {\bibfnamefont {G.~G.}\ \bibnamefont
  {{Howes}}}, \bibinfo {author} {\bibfnamefont {J.~D.}\ \bibnamefont
  {{Rhudy}}}, \bibinfo {author} {\bibfnamefont {S.~K.}\ \bibnamefont
  {{Terry}}}, \bibinfo {author} {\bibfnamefont {T.~A.}\ \bibnamefont
  {{Carter}}}, \bibinfo {author} {\bibfnamefont {C.~A.}\ \bibnamefont
  {{Kletzing}}}, \bibinfo {author} {\bibfnamefont {J.~W.~R.}\ \bibnamefont
  {{Schroeder}}}, \ and\ \bibinfo {author} {\bibfnamefont {F.}~\bibnamefont
  {{Skiff}}},\ }\href {\doibase 10.1063/1.4941977} {\bibfield  {journal}
  {\bibinfo  {journal} {\pop}\ }\textbf {\bibinfo {volume} {23}},\ \bibinfo
  {eid} {022305} (\bibinfo {year} {2016})}\BibitemShut {NoStop}%
\bibitem [{\citenamefont {{Ren}}\ \emph {et~al.}(2011)\citenamefont {{Ren}},
  \citenamefont {{Almagri}}, \citenamefont {{Fiksel}}, \citenamefont
  {{Prager}}, \citenamefont {{Sarff}},\ and\ \citenamefont {{Terry}}}]{ren11}%
  \BibitemOpen
  \bibfield  {author} {\bibinfo {author} {\bibfnamefont {Y.}~\bibnamefont
  {{Ren}}}, \bibinfo {author} {\bibfnamefont {A.~F.}\ \bibnamefont
  {{Almagri}}}, \bibinfo {author} {\bibfnamefont {G.}~\bibnamefont {{Fiksel}}},
  \bibinfo {author} {\bibfnamefont {S.~C.}\ \bibnamefont {{Prager}}}, \bibinfo
  {author} {\bibfnamefont {J.~S.}\ \bibnamefont {{Sarff}}}, \ and\ \bibinfo
  {author} {\bibfnamefont {P.~W.}\ \bibnamefont {{Terry}}},\ }\href {\doibase
  10.1103/PhysRevLett.107.195002} {\bibfield  {journal} {\bibinfo  {journal}
  {\prl}\ }\textbf {\bibinfo {volume} {107}},\ \bibinfo {eid} {195002}
  (\bibinfo {year} {2011})}\BibitemShut {NoStop}%
\bibitem [{\citenamefont {{Thuecks}}\ \emph {et~al.}(2017)\citenamefont
  {{Thuecks}}, \citenamefont {{Almagri}}, \citenamefont {{Sarff}},\ and\
  \citenamefont {{Terry}}}]{thuecks17}%
  \BibitemOpen
  \bibfield  {author} {\bibinfo {author} {\bibfnamefont {D.~J.}\ \bibnamefont
  {{Thuecks}}}, \bibinfo {author} {\bibfnamefont {A.~F.}\ \bibnamefont
  {{Almagri}}}, \bibinfo {author} {\bibfnamefont {J.~S.}\ \bibnamefont
  {{Sarff}}}, \ and\ \bibinfo {author} {\bibfnamefont {P.~W.}\ \bibnamefont
  {{Terry}}},\ }\href {\doibase 10.1063/1.4976838} {\bibfield  {journal}
  {\bibinfo  {journal} {\pop}\ }\textbf {\bibinfo {volume} {24}},\ \bibinfo
  {eid} {022309} (\bibinfo {year} {2017})}\BibitemShut {NoStop}%
\bibitem [{\citenamefont {{Schaffner}}\ \emph
  {et~al.}(2014{\natexlab{a}})\citenamefont {{Schaffner}}, \citenamefont
  {{Wan}},\ and\ \citenamefont {{Brown}}}]{schaffner14a}%
  \BibitemOpen
  \bibfield  {author} {\bibinfo {author} {\bibfnamefont {D.~A.}\ \bibnamefont
  {{Schaffner}}}, \bibinfo {author} {\bibfnamefont {A.}~\bibnamefont {{Wan}}},
  \ and\ \bibinfo {author} {\bibfnamefont {M.~R.}\ \bibnamefont {{Brown}}},\
  }\href {\doibase 10.1103/PhysRevLett.112.165001} {\bibfield  {journal}
  {\bibinfo  {journal} {\prl}\ }\textbf {\bibinfo {volume} {112}},\ \bibinfo
  {eid} {165001} (\bibinfo {year} {2014}{\natexlab{a}})}\BibitemShut {NoStop}%
\bibitem [{\citenamefont {{Schaffner}}\ \emph
  {et~al.}(2014{\natexlab{b}})\citenamefont {{Schaffner}}, \citenamefont
  {{Brown}},\ and\ \citenamefont {{Lukin}}}]{schaffner14b}%
  \BibitemOpen
  \bibfield  {author} {\bibinfo {author} {\bibfnamefont {D.~A.}\ \bibnamefont
  {{Schaffner}}}, \bibinfo {author} {\bibfnamefont {M.~R.}\ \bibnamefont
  {{Brown}}}, \ and\ \bibinfo {author} {\bibfnamefont {V.~S.}\ \bibnamefont
  {{Lukin}}},\ }\href {\doibase 10.1088/0004-637X/790/2/126} {\bibfield
  {journal} {\bibinfo  {journal} {\apj}\ }\textbf {\bibinfo {volume} {790}},\
  \bibinfo {eid} {126} (\bibinfo {year} {2014}{\natexlab{b}})}\BibitemShut
  {NoStop}%
\bibitem [{\citenamefont {{Schaffner}}\ \emph
  {et~al.}(2014{\natexlab{c}})\citenamefont {{Schaffner}}, \citenamefont
  {{Lukin}}, \citenamefont {{Wan}},\ and\ \citenamefont
  {{Brown}}}]{schaffner14c}%
  \BibitemOpen
  \bibfield  {author} {\bibinfo {author} {\bibfnamefont {D.~A.}\ \bibnamefont
  {{Schaffner}}}, \bibinfo {author} {\bibfnamefont {V.~S.}\ \bibnamefont
  {{Lukin}}}, \bibinfo {author} {\bibfnamefont {A.}~\bibnamefont {{Wan}}}, \
  and\ \bibinfo {author} {\bibfnamefont {M.~R.}\ \bibnamefont {{Brown}}},\
  }\href {\doibase 10.1088/0741-3335/56/6/064003} {\bibfield  {journal}
  {\bibinfo  {journal} {\ppcf}\ }\textbf {\bibinfo {volume} {56}},\ \bibinfo
  {eid} {064003} (\bibinfo {year} {2014}{\natexlab{c}})}\BibitemShut {NoStop}%
\bibitem [{\citenamefont {{Schaffner}}\ and\ \citenamefont
  {{Brown}}(2015)}]{schaffner15}%
  \BibitemOpen
  \bibfield  {author} {\bibinfo {author} {\bibfnamefont {D.~A.}\ \bibnamefont
  {{Schaffner}}}\ and\ \bibinfo {author} {\bibfnamefont {M.~R.}\ \bibnamefont
  {{Brown}}},\ }\href {\doibase 10.1088/0004-637X/811/1/61} {\bibfield
  {journal} {\bibinfo  {journal} {\apj}\ }\textbf {\bibinfo {volume} {811}},\
  \bibinfo {eid} {61} (\bibinfo {year} {2015})}\BibitemShut {NoStop}%
\bibitem [{\citenamefont {{Chatterjee}}\ \emph {et~al.}(2017)\citenamefont
  {{Chatterjee}}, \citenamefont {{Schoeffler}}, \citenamefont {{Kumar Singh}},
  \citenamefont {{Adak}}, \citenamefont {{Lad}}, \citenamefont {{Sengupta}},
  \citenamefont {{Kaw}}, \citenamefont {{Silva}}, \citenamefont {{Das}},\ and\
  \citenamefont {{Kumar}}}]{chatterjee17}%
  \BibitemOpen
  \bibfield  {author} {\bibinfo {author} {\bibfnamefont {G.}~\bibnamefont
  {{Chatterjee}}}, \bibinfo {author} {\bibfnamefont {K.~M.}\ \bibnamefont
  {{Schoeffler}}}, \bibinfo {author} {\bibfnamefont {P.}~\bibnamefont {{Kumar
  Singh}}}, \bibinfo {author} {\bibfnamefont {A.}~\bibnamefont {{Adak}}},
  \bibinfo {author} {\bibfnamefont {A.~D.}\ \bibnamefont {{Lad}}}, \bibinfo
  {author} {\bibfnamefont {S.}~\bibnamefont {{Sengupta}}}, \bibinfo {author}
  {\bibfnamefont {P.}~\bibnamefont {{Kaw}}}, \bibinfo {author} {\bibfnamefont
  {L.~O.}\ \bibnamefont {{Silva}}}, \bibinfo {author} {\bibfnamefont
  {A.}~\bibnamefont {{Das}}}, \ and\ \bibinfo {author} {\bibfnamefont {G.~R.}\
  \bibnamefont {{Kumar}}},\ }\href {\doibase 10.1038/ncomms15970} {\bibfield
  {journal} {\bibinfo  {journal} {\natcomm}\ }\textbf {\bibinfo {volume} {8}},\
  \bibinfo {eid} {15970} (\bibinfo {year} {2017})}\BibitemShut {NoStop}%
\bibitem [{\citenamefont {{Richner}}\ \emph {et~al.}(2022)\citenamefont
  {{Richner}}, \citenamefont {{Bodner}}, \citenamefont {{Bongard}},
  \citenamefont {{Fonck}}, \citenamefont {{Nornberg}},\ and\ \citenamefont
  {{Reusch}}}]{richner22}%
  \BibitemOpen
  \bibfield  {author} {\bibinfo {author} {\bibfnamefont {N.~J.}\ \bibnamefont
  {{Richner}}}, \bibinfo {author} {\bibfnamefont {G.~M.}\ \bibnamefont
  {{Bodner}}}, \bibinfo {author} {\bibfnamefont {M.~W.}\ \bibnamefont
  {{Bongard}}}, \bibinfo {author} {\bibfnamefont {R.~J.}\ \bibnamefont
  {{Fonck}}}, \bibinfo {author} {\bibfnamefont {M.~D.}\ \bibnamefont
  {{Nornberg}}}, \ and\ \bibinfo {author} {\bibfnamefont {J.~A.}\ \bibnamefont
  {{Reusch}}},\ }\href {\doibase 10.1103/PhysRevLett.128.105001} {\bibfield
  {journal} {\bibinfo  {journal} {\prl}\ }\textbf {\bibinfo {volume} {128}},\
  \bibinfo {eid} {105001} (\bibinfo {year} {2022})}\BibitemShut {NoStop}%
\bibitem [{\citenamefont {{Gekelman}}\ \emph {et~al.}(2016)\citenamefont
  {{Gekelman}}, \citenamefont {{Pribyl}}, \citenamefont {{Lucky}},
  \citenamefont {{Drandell}}, \citenamefont {{Leneman}}, \citenamefont
  {{Maggs}}, \citenamefont {{Vincena}}, \citenamefont {{Van Compernolle}},
  \citenamefont {{Tripathi}}, \citenamefont {{Morales}}, \citenamefont
  {{Carter}}, \citenamefont {{Wang}},\ and\ \citenamefont
  {{DeHaas}}}]{gekelman16}%
  \BibitemOpen
  \bibfield  {author} {\bibinfo {author} {\bibfnamefont {W.}~\bibnamefont
  {{Gekelman}}}, \bibinfo {author} {\bibfnamefont {P.}~\bibnamefont
  {{Pribyl}}}, \bibinfo {author} {\bibfnamefont {Z.}~\bibnamefont {{Lucky}}},
  \bibinfo {author} {\bibfnamefont {M.}~\bibnamefont {{Drandell}}}, \bibinfo
  {author} {\bibfnamefont {D.}~\bibnamefont {{Leneman}}}, \bibinfo {author}
  {\bibfnamefont {J.}~\bibnamefont {{Maggs}}}, \bibinfo {author} {\bibfnamefont
  {S.}~\bibnamefont {{Vincena}}}, \bibinfo {author} {\bibfnamefont
  {B.}~\bibnamefont {{Van Compernolle}}}, \bibinfo {author} {\bibfnamefont
  {S.~K.~P.}\ \bibnamefont {{Tripathi}}}, \bibinfo {author} {\bibfnamefont
  {G.}~\bibnamefont {{Morales}}}, \bibinfo {author} {\bibfnamefont {T.~A.}\
  \bibnamefont {{Carter}}}, \bibinfo {author} {\bibfnamefont {Y.}~\bibnamefont
  {{Wang}}}, \ and\ \bibinfo {author} {\bibfnamefont {T.}~\bibnamefont
  {{DeHaas}}},\ }\href {\doibase 10.1063/1.4941079} {\bibfield  {journal}
  {\bibinfo  {journal} {\rsi}\ }\textbf {\bibinfo {volume} {87}},\ \bibinfo
  {eid} {025105} (\bibinfo {year} {2016})}\BibitemShut {NoStop}%
\bibitem [{\citenamefont {{Leneman}}\ \emph {et~al.}(2006)\citenamefont
  {{Leneman}}, \citenamefont {{Gekelman}},\ and\ \citenamefont
  {{Maggs}}}]{leneman06}%
  \BibitemOpen
  \bibfield  {author} {\bibinfo {author} {\bibfnamefont {D.}~\bibnamefont
  {{Leneman}}}, \bibinfo {author} {\bibfnamefont {W.}~\bibnamefont
  {{Gekelman}}}, \ and\ \bibinfo {author} {\bibfnamefont {J.}~\bibnamefont
  {{Maggs}}},\ }\href {\doibase 10.1063/1.2150829} {\bibfield  {journal}
  {\bibinfo  {journal} {\rsi}\ }\textbf {\bibinfo {volume} {77}},\ \bibinfo
  {eid} {015108} (\bibinfo {year} {2006})}\BibitemShut {NoStop}%
\bibitem [{\citenamefont {{Gigliotti}}\ \emph {et~al.}(2009)\citenamefont
  {{Gigliotti}}, \citenamefont {{Gekelman}}, \citenamefont {{Pribyl}},
  \citenamefont {{Vincena}}, \citenamefont {{Karavaev}}, \citenamefont
  {{Shao}}, \citenamefont {{Sharma}},\ and\ \citenamefont
  {{Papadopoulos}}}]{gigliotti09}%
  \BibitemOpen
  \bibfield  {author} {\bibinfo {author} {\bibfnamefont {A.}~\bibnamefont
  {{Gigliotti}}}, \bibinfo {author} {\bibfnamefont {W.}~\bibnamefont
  {{Gekelman}}}, \bibinfo {author} {\bibfnamefont {P.}~\bibnamefont
  {{Pribyl}}}, \bibinfo {author} {\bibfnamefont {S.}~\bibnamefont {{Vincena}}},
  \bibinfo {author} {\bibfnamefont {A.}~\bibnamefont {{Karavaev}}}, \bibinfo
  {author} {\bibfnamefont {X.}~\bibnamefont {{Shao}}}, \bibinfo {author}
  {\bibfnamefont {A.~S.}\ \bibnamefont {{Sharma}}}, \ and\ \bibinfo {author}
  {\bibfnamefont {D.}~\bibnamefont {{Papadopoulos}}},\ }\href {\doibase
  10.1063/1.3224030} {\bibfield  {journal} {\bibinfo  {journal} {\pop}\
  }\textbf {\bibinfo {volume} {16}},\ \bibinfo {eid} {092106} (\bibinfo {year}
  {2009})}\BibitemShut {NoStop}%
\bibitem [{\citenamefont {{Bose}}\ \emph {et~al.}(2019)\citenamefont {{Bose}},
  \citenamefont {{Kaur}}, \citenamefont {{Barada}}, \citenamefont {{Ghosh}},
  \citenamefont {{Chattopadhyay}},\ and\ \citenamefont {{Pal}}}]{bose19b}%
  \BibitemOpen
  \bibfield  {author} {\bibinfo {author} {\bibfnamefont {S.}~\bibnamefont
  {{Bose}}}, \bibinfo {author} {\bibfnamefont {M.}~\bibnamefont {{Kaur}}},
  \bibinfo {author} {\bibfnamefont {K.~K.}\ \bibnamefont {{Barada}}}, \bibinfo
  {author} {\bibfnamefont {J.}~\bibnamefont {{Ghosh}}}, \bibinfo {author}
  {\bibfnamefont {P.~K.}\ \bibnamefont {{Chattopadhyay}}}, \ and\ \bibinfo
  {author} {\bibfnamefont {R.}~\bibnamefont {{Pal}}},\ }\href {\doibase
  10.1088/1361-6404/aaee31} {\bibfield  {journal} {\bibinfo  {journal} {\ejp}\
  }\textbf {\bibinfo {volume} {40}},\ \bibinfo {pages} {015803} (\bibinfo
  {year} {2019})}\BibitemShut {NoStop}%
\bibitem [{\citenamefont {{Torrence}}\ and\ \citenamefont
  {{Compo}}(1998)}]{torrence98}%
  \BibitemOpen
  \bibfield  {author} {\bibinfo {author} {\bibfnamefont {C.}~\bibnamefont
  {{Torrence}}}\ and\ \bibinfo {author} {\bibfnamefont {G.~P.}\ \bibnamefont
  {{Compo}}},\ }\href {\doibase
  10.1175/1520-0477(1998)079<0061:APGTWA>2.0.CO;2} {\bibfield  {journal}
  {\bibinfo  {journal} {\bmas}\ }\textbf {\bibinfo {volume} {79}},\ \bibinfo
  {pages} {61} (\bibinfo {year} {1998})}\BibitemShut {NoStop}%
\bibitem [{\citenamefont {{Hollweg}}(1999)}]{hollweg99}%
  \BibitemOpen
  \bibfield  {author} {\bibinfo {author} {\bibfnamefont {J.~V.}\ \bibnamefont
  {{Hollweg}}},\ }\href {\doibase 10.1029/1998JA900132} {\bibfield  {journal}
  {\bibinfo  {journal} {\jgr}\ }\textbf {\bibinfo {volume} {104}},\ \bibinfo
  {pages} {14811} (\bibinfo {year} {1999})}\BibitemShut {NoStop}%
\bibitem [{\citenamefont {{Chen}}\ \emph {et~al.}(2014)\citenamefont {{Chen}},
  \citenamefont {{Leung}}, \citenamefont {{Boldyrev}}, \citenamefont
  {{Maruca}},\ and\ \citenamefont {{Bale}}}]{chen14b}%
  \BibitemOpen
  \bibfield  {author} {\bibinfo {author} {\bibfnamefont {C.~H.~K.}\
  \bibnamefont {{Chen}}}, \bibinfo {author} {\bibfnamefont {L.}~\bibnamefont
  {{Leung}}}, \bibinfo {author} {\bibfnamefont {S.}~\bibnamefont {{Boldyrev}}},
  \bibinfo {author} {\bibfnamefont {B.~A.}\ \bibnamefont {{Maruca}}}, \ and\
  \bibinfo {author} {\bibfnamefont {S.~D.}\ \bibnamefont {{Bale}}},\ }\href
  {\doibase 10.1002/2014GL062009} {\bibfield  {journal} {\bibinfo  {journal}
  {\grl}\ }\textbf {\bibinfo {volume} {41}},\ \bibinfo {pages} {8081} (\bibinfo
  {year} {2014})}\BibitemShut {NoStop}%
\bibitem [{\citenamefont {{Boldyrev}}\ \emph {et~al.}(2015)\citenamefont
  {{Boldyrev}}, \citenamefont {{Chen}}, \citenamefont {{Xia}},\ and\
  \citenamefont {{Zhdankin}}}]{boldyrev15}%
  \BibitemOpen
  \bibfield  {author} {\bibinfo {author} {\bibfnamefont {S.}~\bibnamefont
  {{Boldyrev}}}, \bibinfo {author} {\bibfnamefont {C.~H.~K.}\ \bibnamefont
  {{Chen}}}, \bibinfo {author} {\bibfnamefont {Q.}~\bibnamefont {{Xia}}}, \
  and\ \bibinfo {author} {\bibfnamefont {V.}~\bibnamefont {{Zhdankin}}},\
  }\href {\doibase 10.1088/0004-637X/806/2/238} {\bibfield  {journal} {\bibinfo
   {journal} {\apj}\ }\textbf {\bibinfo {volume} {806}},\ \bibinfo {eid} {238}
  (\bibinfo {year} {2015})}\BibitemShut {NoStop}%
\bibitem [{\citenamefont {{Mallet}}\ \emph {et~al.}(2023)\citenamefont
  {{Mallet}}, \citenamefont {{Dorfman}}, \citenamefont {{Abler}}, \citenamefont
  {{Bowen}},\ and\ \citenamefont {{Chen}}}]{mallet23}%
  \BibitemOpen
  \bibfield  {author} {\bibinfo {author} {\bibfnamefont {A.}~\bibnamefont
  {{Mallet}}}, \bibinfo {author} {\bibfnamefont {S.}~\bibnamefont {{Dorfman}}},
  \bibinfo {author} {\bibfnamefont {M.}~\bibnamefont {{Abler}}}, \bibinfo
  {author} {\bibfnamefont {T.~A.}\ \bibnamefont {{Bowen}}}, \ and\ \bibinfo
  {author} {\bibfnamefont {C.~H.~K.}\ \bibnamefont {{Chen}}},\ }\href {\doibase
  10.1063/5.0151035} {\bibfield  {journal} {\bibinfo  {journal} {\pop}\
  }\textbf {\bibinfo {volume} {30}},\ \bibinfo {eid} {112102} (\bibinfo {year}
  {2023})}\BibitemShut {NoStop}%
\bibitem [{\citenamefont {{Strauss}}(1976)}]{strauss76}%
  \BibitemOpen
  \bibfield  {author} {\bibinfo {author} {\bibfnamefont {H.~R.}\ \bibnamefont
  {{Strauss}}},\ }\href {\doibase 10.1063/1.861310} {\bibfield  {journal}
  {\bibinfo  {journal} {\pof}\ }\textbf {\bibinfo {volume} {19}},\ \bibinfo
  {pages} {134} (\bibinfo {year} {1976})}\BibitemShut {NoStop}%
\bibitem [{\citenamefont {{Schekochihin}}\ \emph {et~al.}(2009)\citenamefont
  {{Schekochihin}}, \citenamefont {{Cowley}}, \citenamefont {{Dorland}},
  \citenamefont {{Hammett}}, \citenamefont {{Howes}}, \citenamefont
  {{Quataert}},\ and\ \citenamefont {{Tatsuno}}}]{schekochihin09}%
  \BibitemOpen
  \bibfield  {author} {\bibinfo {author} {\bibfnamefont {A.~A.}\ \bibnamefont
  {{Schekochihin}}}, \bibinfo {author} {\bibfnamefont {S.~C.}\ \bibnamefont
  {{Cowley}}}, \bibinfo {author} {\bibfnamefont {W.}~\bibnamefont {{Dorland}}},
  \bibinfo {author} {\bibfnamefont {G.~W.}\ \bibnamefont {{Hammett}}}, \bibinfo
  {author} {\bibfnamefont {G.~G.}\ \bibnamefont {{Howes}}}, \bibinfo {author}
  {\bibfnamefont {E.}~\bibnamefont {{Quataert}}}, \ and\ \bibinfo {author}
  {\bibfnamefont {T.}~\bibnamefont {{Tatsuno}}},\ }\href {\doibase
  10.1088/0067-0049/182/1/310} {\bibfield  {journal} {\bibinfo  {journal}
  {\apjs}\ }\textbf {\bibinfo {volume} {182}},\ \bibinfo {pages} {310}
  (\bibinfo {year} {2009})}\BibitemShut {NoStop}%
\bibitem [{\citenamefont {{Franci}}\ \emph {et~al.}(2018)\citenamefont
  {{Franci}}, \citenamefont {{Hellinger}}, \citenamefont {{Guarrasi}},
  \citenamefont {{Chen}}, \citenamefont {{Papini}}, \citenamefont {{Verdini}},
  \citenamefont {{Matteini}},\ and\ \citenamefont {{Landi}}}]{franci18}%
  \BibitemOpen
  \bibfield  {author} {\bibinfo {author} {\bibfnamefont {L.}~\bibnamefont
  {{Franci}}}, \bibinfo {author} {\bibfnamefont {P.}~\bibnamefont
  {{Hellinger}}}, \bibinfo {author} {\bibfnamefont {M.}~\bibnamefont
  {{Guarrasi}}}, \bibinfo {author} {\bibfnamefont {C.~H.~K.}\ \bibnamefont
  {{Chen}}}, \bibinfo {author} {\bibfnamefont {E.}~\bibnamefont {{Papini}}},
  \bibinfo {author} {\bibfnamefont {A.}~\bibnamefont {{Verdini}}}, \bibinfo
  {author} {\bibfnamefont {L.}~\bibnamefont {{Matteini}}}, \ and\ \bibinfo
  {author} {\bibfnamefont {S.}~\bibnamefont {{Landi}}},\ }in\ \href {\doibase
  10.1088/1742-6596/1031/1/012002} {\emph {\bibinfo {booktitle} {Journal of
  Physics Conference Series}}},\ \bibinfo {series} {Journal of Physics
  Conference Series}, Vol.\ \bibinfo {volume} {1031}\ (\bibinfo  {publisher}
  {IOP},\ \bibinfo {year} {2018})\ p.\ \bibinfo {pages} {012002}\BibitemShut
  {NoStop}%
\bibitem [{\citenamefont {{Voitenko}}(1998)}]{voitenko98}%
  \BibitemOpen
  \bibfield  {author} {\bibinfo {author} {\bibfnamefont {Y.~M.}\ \bibnamefont
  {{Voitenko}}},\ }\href {\doibase 10.1017/S0022377898007107} {\bibfield
  {journal} {\bibinfo  {journal} {\jpp}\ }\textbf {\bibinfo {volume} {60}},\
  \bibinfo {pages} {515} (\bibinfo {year} {1998})}\BibitemShut {NoStop}%
\bibitem [{\citenamefont {{Cho}}(2011)}]{cho11}%
  \BibitemOpen
  \bibfield  {author} {\bibinfo {author} {\bibfnamefont {J.}~\bibnamefont
  {{Cho}}},\ }\href {\doibase 10.1103/PhysRevLett.106.191104} {\bibfield
  {journal} {\bibinfo  {journal} {\prl}\ }\textbf {\bibinfo {volume} {106}},\
  \bibinfo {eid} {191104} (\bibinfo {year} {2011})}\BibitemShut {NoStop}%
\bibitem [{\citenamefont {{Voitenko}}\ and\ \citenamefont {{de
  Keyser}}(2011)}]{voitenko11}%
  \BibitemOpen
  \bibfield  {author} {\bibinfo {author} {\bibfnamefont {Y.}~\bibnamefont
  {{Voitenko}}}\ and\ \bibinfo {author} {\bibfnamefont {J.}~\bibnamefont {{de
  Keyser}}},\ }\href {\doibase 10.5194/npg-18-587-2011} {\bibfield  {journal}
  {\bibinfo  {journal} {\npg}\ }\textbf {\bibinfo {volume} {18}},\ \bibinfo
  {pages} {587} (\bibinfo {year} {2011})}\BibitemShut {NoStop}%
\bibitem [{\citenamefont {{Voitenko}}\ and\ \citenamefont {{De
  Keyser}}(2016)}]{voitenko16}%
  \BibitemOpen
  \bibfield  {author} {\bibinfo {author} {\bibfnamefont {Y.}~\bibnamefont
  {{Voitenko}}}\ and\ \bibinfo {author} {\bibfnamefont {J.}~\bibnamefont {{De
  Keyser}}},\ }\href {\doibase 10.3847/2041-8205/832/2/L20} {\bibfield
  {journal} {\bibinfo  {journal} {\apjl}\ }\textbf {\bibinfo {volume} {832}},\
  \bibinfo {eid} {L20} (\bibinfo {year} {2016})}\BibitemShut {NoStop}%
\bibitem [{\citenamefont {{Chen}}\ \emph {et~al.}(2020)\citenamefont {{Chen}},
  \citenamefont {{Bale}}, \citenamefont {{Bonnell}}, \citenamefont
  {{Borovikov}}, \citenamefont {{Bowen}}, \citenamefont {{Burgess}},
  \citenamefont {{Case}}, \citenamefont {{Chandran}}, \citenamefont {{de Wit}},
  \citenamefont {{Goetz}}, \citenamefont {{Harvey}}, \citenamefont {{Kasper}},
  \citenamefont {{Klein}}, \citenamefont {{Korreck}}, \citenamefont {{Larson}},
  \citenamefont {{Livi}}, \citenamefont {{MacDowall}}, \citenamefont
  {{Malaspina}}, \citenamefont {{Mallet}}, \citenamefont {{McManus}},
  \citenamefont {{Moncuquet}}, \citenamefont {{Pulupa}}, \citenamefont
  {{Stevens}},\ and\ \citenamefont {{Whittlesey}}}]{chen20}%
  \BibitemOpen
  \bibfield  {author} {\bibinfo {author} {\bibfnamefont {C.~H.~K.}\
  \bibnamefont {{Chen}}}, \bibinfo {author} {\bibfnamefont {S.~D.}\
  \bibnamefont {{Bale}}}, \bibinfo {author} {\bibfnamefont {J.~W.}\
  \bibnamefont {{Bonnell}}}, \bibinfo {author} {\bibfnamefont {D.}~\bibnamefont
  {{Borovikov}}}, \bibinfo {author} {\bibfnamefont {T.~A.}\ \bibnamefont
  {{Bowen}}}, \bibinfo {author} {\bibfnamefont {D.}~\bibnamefont {{Burgess}}},
  \bibinfo {author} {\bibfnamefont {A.~W.}\ \bibnamefont {{Case}}}, \bibinfo
  {author} {\bibfnamefont {B.~D.~G.}\ \bibnamefont {{Chandran}}}, \bibinfo
  {author} {\bibfnamefont {T.~D.}\ \bibnamefont {{de Wit}}}, \bibinfo {author}
  {\bibfnamefont {K.}~\bibnamefont {{Goetz}}}, \bibinfo {author} {\bibfnamefont
  {P.~R.}\ \bibnamefont {{Harvey}}}, \bibinfo {author} {\bibfnamefont {J.~C.}\
  \bibnamefont {{Kasper}}}, \bibinfo {author} {\bibfnamefont {K.~G.}\
  \bibnamefont {{Klein}}}, \bibinfo {author} {\bibfnamefont {K.~E.}\
  \bibnamefont {{Korreck}}}, \bibinfo {author} {\bibfnamefont {D.}~\bibnamefont
  {{Larson}}}, \bibinfo {author} {\bibfnamefont {R.}~\bibnamefont {{Livi}}},
  \bibinfo {author} {\bibfnamefont {R.~J.}\ \bibnamefont {{MacDowall}}},
  \bibinfo {author} {\bibfnamefont {D.~M.}\ \bibnamefont {{Malaspina}}},
  \bibinfo {author} {\bibfnamefont {A.}~\bibnamefont {{Mallet}}}, \bibinfo
  {author} {\bibfnamefont {M.~D.}\ \bibnamefont {{McManus}}}, \bibinfo {author}
  {\bibfnamefont {M.}~\bibnamefont {{Moncuquet}}}, \bibinfo {author}
  {\bibfnamefont {M.}~\bibnamefont {{Pulupa}}}, \bibinfo {author}
  {\bibfnamefont {M.~L.}\ \bibnamefont {{Stevens}}}, \ and\ \bibinfo {author}
  {\bibfnamefont {P.}~\bibnamefont {{Whittlesey}}},\ }\href {\doibase
  10.3847/1538-4365/ab60a3} {\bibfield  {journal} {\bibinfo  {journal} {\apjs}\
  }\textbf {\bibinfo {volume} {246}},\ \bibinfo {eid} {53} (\bibinfo {year}
  {2020})}\BibitemShut {NoStop}%
\bibitem [{\citenamefont {{Meyrand}}\ \emph {et~al.}(2021)\citenamefont
  {{Meyrand}}, \citenamefont {{Squire}}, \citenamefont {{Schekochihin}},\ and\
  \citenamefont {{Dorland}}}]{meyrand21}%
  \BibitemOpen
  \bibfield  {author} {\bibinfo {author} {\bibfnamefont {R.}~\bibnamefont
  {{Meyrand}}}, \bibinfo {author} {\bibfnamefont {J.}~\bibnamefont {{Squire}}},
  \bibinfo {author} {\bibfnamefont {A.~A.}\ \bibnamefont {{Schekochihin}}}, \
  and\ \bibinfo {author} {\bibfnamefont {W.}~\bibnamefont {{Dorland}}},\ }\href
  {\doibase 10.1017/S0022377821000489} {\bibfield  {journal} {\bibinfo
  {journal} {\jpp}\ }\textbf {\bibinfo {volume} {87}},\ \bibinfo {eid}
  {535870301} (\bibinfo {year} {2021})}\BibitemShut {NoStop}%
\bibitem [{\citenamefont {{Squire}}\ \emph {et~al.}(2022)\citenamefont
  {{Squire}}, \citenamefont {{Meyrand}}, \citenamefont {{Kunz}}, \citenamefont
  {{Arzamasskiy}}, \citenamefont {{Schekochihin}},\ and\ \citenamefont
  {{Quataert}}}]{squire22}%
  \BibitemOpen
  \bibfield  {author} {\bibinfo {author} {\bibfnamefont {J.}~\bibnamefont
  {{Squire}}}, \bibinfo {author} {\bibfnamefont {R.}~\bibnamefont {{Meyrand}}},
  \bibinfo {author} {\bibfnamefont {M.~W.}\ \bibnamefont {{Kunz}}}, \bibinfo
  {author} {\bibfnamefont {L.}~\bibnamefont {{Arzamasskiy}}}, \bibinfo {author}
  {\bibfnamefont {A.~A.}\ \bibnamefont {{Schekochihin}}}, \ and\ \bibinfo
  {author} {\bibfnamefont {E.}~\bibnamefont {{Quataert}}},\ }\href@noop {}
  {\bibfield  {journal} {\bibinfo  {journal} {Nat. Astron.}\ } (\bibinfo {year}
  {2022})}\BibitemShut {NoStop}%
\bibitem [{\citenamefont {{McIntyre}}\ \emph {et~al.}(2025)\citenamefont
  {{McIntyre}}, \citenamefont {{Chen}}, \citenamefont {{Squire}}, \citenamefont
  {{Meyrand}},\ and\ \citenamefont {{Simon}}}]{mcintyre25}%
  \BibitemOpen
  \bibfield  {author} {\bibinfo {author} {\bibfnamefont {J.~R.}\ \bibnamefont
  {{McIntyre}}}, \bibinfo {author} {\bibfnamefont {C.~H.~K.}\ \bibnamefont
  {{Chen}}}, \bibinfo {author} {\bibfnamefont {J.}~\bibnamefont {{Squire}}},
  \bibinfo {author} {\bibfnamefont {R.}~\bibnamefont {{Meyrand}}}, \ and\
  \bibinfo {author} {\bibfnamefont {P.~A.}\ \bibnamefont {{Simon}}},\ }\href
  {\doibase 10.1103/PhysRevX.15.031008} {\bibfield  {journal} {\bibinfo
  {journal} {\prx}\ }\textbf {\bibinfo {volume} {15}},\ \bibinfo {eid} {031008}
  (\bibinfo {year} {2025})}\BibitemShut {NoStop}%
\end{thebibliography}%

\end{document}